*Jan 15, 2025_Version 2 (originally submitted on Dec 21, 2024_Version 1)*

# A unified model for the origins of spongiform degeneration and other neuropathological features in prion diseases

Gerold Schmitt-Ulms[1,2*], Xinzhu Wang[1,2], Joel Watts[1,3], Stephanie Booth[4], Holger Wille[5,6], Wenda Zhao[1,2]

[1]Tanz Centre for Research in Neurodegenerative Diseases, University of Toronto, Toronto, Canada.
[2]Department of Laboratory Medicine & Pathobiology, University of Toronto, Toronto, Canada.

[3]Department of Biochemistry, University of Toronto, Toronto, Canada.

[4]Department of Medical Microbiology and Infectious Diseases, University of Manitoba, Winnipeg, Canada.

[5]Department of Biochemistry, University of Alberta, Edmonton, Canada.

[6]Centre for Prions and Protein Folding Diseases, University of Edmonton, Edmonton, Canada.

* Please address correspondence to: g.schmittulms@utoronto.ca.

Running title: A unified model of neuropathological features in prion diseases

## Abstract

Decades after their initial observation in prion-infected brain tissues, the identities of virus-like dense particles, varicose tubules, and oval bodies containing parallel bands and fibrils have remained elusive. Our recent work revealed that a phenotype of dilation of the endoplasmic reticulum (ER), most notable for the perinuclear space (PNS), contributes to spongiform degeneration. To assess the significance of this phenotype for the etiology of prion diseases, we explored whether it can be functionally linked to other neuropathological hallmarks observed in these diseases, as this would indicate it to be a central event. Having surveyed the neuropathological record and other distant literature niches, we propose a model in which pathogenic forms of the prion protein poison raft domains, including essential $Na^+$, $K^+$-ATPases (NKAs) embedded within them, thereby triggering an ER-centered cellular rescue program coordinated by the unfolded protein response (UPR). The execution of this program stalls general protein synthesis, causing the deterioration of synaptic spines. As the disease progresses, cells selectively increase sterol biosynthesis, along with ribosome and ER biogenesis. These adaptive rescue attempts cause morphological changes to the ER which manifest as ER dilation or ER hypertrophy in a manner that is influenced by $Ca^{2+}$ influx into the cell. The nuclear-to-cytoplasmic transport of mRNAs and tRNAs interrupts in late stage disease, thereby depriving ribosomes of supplies and inducing them to aggregate into a paracrystalline form. In support of this model, we share previously reported data, whose features are consistent with the interpretation that 1) the phenotype of ER dilation is observed in major prion diseases, 2) varicose tubules and oval bodies represent ER hypertrophy, and 3) virus-like dense particles are paracrystalline aggregates of inactive ribosomes.

## Introduction

The appearance of sponge-like (spongiform) degeneration in the brain parenchyma is a hallmark of a group of fatal neurodegenerative diseases known as prion diseases. Both the identity of the subcellular structures that give rise to vacuoles in these diseases and the mechanistic basis of their dilation are still debated a hundred years after the original description of the first human cases.

We previously showed that the cellular prion protein ($PrP^C$) interacts with the neural cell adhesion molecule 1 (NCAM1) (1, 2), a protein we deem to be its main functional partner (3, 4), and $Na^+$, $K^+$-ATPases (NKAs) (5). Specifically, we showed that $PrP^C$ resides in raft domains in proximity to NKAs and promotes their activity. NKAs are heterodimeric transmembrane protein complexes, composed of one α-subunit and one β-subunit. Per adenosine triphosphate (ATP) hydrolysis-driven pump cycle, they anti-port two $K^+$ ions inward and three $Na^+$ ions outward. As they do so, they generate the electrochemical gradient, which drives secondary transport processes. It has been estimated that NKAs consume up to three quarters of a neuron's energy expenditure in ATP (6).

Exposure of cells to toxic levels of cardiac glycosides (CGs) leads to an inhibition of NKAs (7, 8). This, in turn, causes a breakdown of the electrochemical gradient (9). When we added increasing doses of CGs to co-cultures of human astrocytes and neurons, we observed steady-state levels of both the NKA and $PrP^C$ to be gradually reduced as we increased non-toxic CG concentrations, yet the cells returned to increasing NKA expression once a toxic CG threshold was reached (10, 11). Next, we documented that certain molecular signatures that can be observed in response to CG exposure also occur in PrP-deficiency or prion disease paradigms (5). These include a profound upregulation of the levels of 5'-nucleotidase, also known as CD73, and the calpain-dependent cleavage of glial fibrillary acidic protein (GFAP).

Consequently, we studied a body of literature that describes the cellular pathology observed in response to CG exposure and became aware of reported similarities to the pathological spongiform degeneration that characterizes prion diseases. In fact, a hypothesis that NKA inhibition may underlie vacuolation in prion diseases was first formulated in 1966 (12) and was further developed in subsequent years (13-16).

A close look at reports, which referenced this early work, revealed that this hypothesis was largely dismissed by others on the grounds that vacuolation in rats exposed to CGs exhibited primarily astrocytic swelling (17, 18), whereas spongiform degeneration in prion diseases is predominantly observed in neurons (19). We hypothesized that this cell type specificity can be accounted for by the relative susceptibility of specific NKA isoforms to CGs: NKA α1 orthologs in rodents, which are expressed in neurons but not in astrocytes, differ from other mammalian α1 subunits in two amino acid positions that render the rodent isoforms refractory to CG inhibition (20, 21). Consistent with our hypothesis, we noticed that CG poisoning had already been shown to induce a vacuolation phenotype that extended to neurons in ouabain-poisoned cats (22). Since then, we have injected toxic CG levels intracerebrally into mice, whose α1 subunit had been rendered CG-sensitive by exchanging the two amino acid positions for the corresponding amino acids present in the human α1 sequence (23). In these *Atp1a1*$^{s/s}$ mice—but not in wild-type control mice of the same genetic background—we observed the ensuing vacuolation to extend to neurons (24).

During these studies, we became aware of a somewhat scattered body of literature describing similar vacuolation phenotypes in a wide variety of paradigms. We hypothesized that the shared morphological abnormalities characterizing this phenotype, particularly the pronounced dilation of the ER—most notably observed in the PNS—result from shared molecular events. Consequently, we surveyed this body of literature with a view to identify common themes.

Since the effects of CG poisoning are relatively well understood, we will initially review in broad strokes cellular responses associated with NKA inhibition. Next, we will highlight similar vacuolation phenotypes in other paradigms. To make sense of them, we grouped them by the type of insult that can lead to this phenotype, most often—yet inconsistently—referred to by the term paraptosis (25). We noticed that the mechanism by which cells increase the turgor in the ER and PNS cisternae, thereby giving rise to vacuolation, has remained murky.

We therefore set out on a protracted experimental program, in its course adapting several previously reported PNS dilation paradigms, with the objective to gain additional mechanistic insights for how large PNS vacuoles form in cells (24). The results pointed to three cellular events that appear central to this phenotype, namely, 1) a characteristic form of ER stress, 2) $Ca^{2+}$ influx, and 3) an upregulation of the sterol biosynthesis along with ribosome and ER biogenesis. Intriguingly, available data suggest that each of these events can also be the initial trigger of this phenotype. Moreover, we propose that the cellular events leading to the ER dilation phenotype are related to those causing a phenomenon known as ER hypertrophy. When we next tried to gauge the potential significance of these phenotypes for the etiology of prion diseases, we noticed that the neuropathological record corroborates the conclusion that ER dilation and ER hypertrophy are distinguishing characteristics of prion diseases. Finally, a search for prior reports on PNS dilations led us to a niche literature on paracrystalline arrangements of inactive ribosomes. The similarities in the size of inactive ribosomes and the arrangements they form in paracrystals to dense particles observed in postmortem prion disease brain sections—sometimes described as ‘virus-like’—suggest to us common origins of these structures.

## 1. NKA inhibition and cellular homeostasis

NKAs drive the physiological electrochemical gradient which is central to secondary transport processes and electric activity in the brain. The electrochemical gradient in vertebrates is characterized by high $Na^+$ and $Cl^-$ concentrations (~100-150 mM) and low $K^+$ levels (~5 mM) in the extracellular space, compared to low $Na^+$ and $Cl^-$ concentrations (~5 mM) and high $K^+$ levels (~100-150 mM) in the cytosol, along with protein-associated ionic charges. The inhibition or disruption of NKAs causes an increase in intracellular $Na^+$ levels, which activates the reverse mode of the $Na^+/Ca^{2+}$ exchanger (NCX), leading to $Ca^{2+}$ influx into the cytosol (26). In electrically active cells, this depolarization is further enhanced by a short chain of events that causes the N-methyl-D-aspartate receptor (NMDA) receptor to import additional $Ca^{2+}$ ions. Others have proposed a direct influence of $PrP^C$ on NMDA activity (27). Our own data favor an indirect effect mediated by excitatory amino acid transporters (EEATs) whose glutamate transport is directly coupled to NKAs (28). Consistent with this account, our recent mouse brain $PrP^C$ interactome analysis documented EAATs—but not NMDA receptor subunits—to co-immunoprecipitate alongside $PrP^C$ and NKAs (5). Consequently, inhibition of NKAs may prompt additional $Ca^{2+}$ influx through the NMDA receptor by reducing glutamate uptake via EEATs, which increases extracellular glutamate levels.

Abnormal increases in intracellular $Ca^{2+}$ levels induce several signaling events, including the activation of calpains and the calcium–calmodulin (CaM) system, along with its associated agents (29). Calpains catalyze the endoproteolysis of specific cellular proteins, including the glial acidic fibrillary protein (GFAP) (30) and the prion protein (31), the latter leading to the production of the C2 cleavage fragment in prion-infected cells (32). As NKAs are the primary consumers of intracellular ATP, any events that diminish pump activity lead to elevated local ATP levels. This, in turn, can stimulate nearby ATP-dependent processes that are typically ATP-limited. It has been proposed that the activating autophosphorylation of Src family kinases (SFKs), a well-document phenomenon observed in proximity to inhibited NKAs (33, 34), is one such event.

Cells respond to insults that compromise or inhibit the pump by removing incapacitated NKAs from the cell surface (10). Because NKAs are embedded in raft domains (35-38), specialized membrane domains rich in cholesterol and sphingolipids, their removal will also deplete the cell surface levels of these raft domains (39). Not surprisingly, along for the ride will also come other proteins enriched in raft domains, including caveolins (40, 41) and $PrP^{C}$ (42). In fact, this passive depletion of $PrP^{C}$ suggested a rational strategy for its removal from the cell surface whose merit as a potential treatment for prion diseases is currently being explored (10, 11, 43). To compensate for the loss of pump activity at the plasma membrane, a cell finding itself in this precarious situation can be expected to invest in the expression of replacement pumps and associated molecules.

To counteract a shortage in raft domains, cells liberate a specialized transcription factor, the sterol regulatory element-binding protein (SREBP) from the ER (44). The latter is accomplished through specific proteolytic processing of SREBP, which precedes an increase in the transcription of genes tasked with sterol biosynthesis (45). The events described above get help from a second mechanism by which reduced levels of NKAs at the plasma membrane can lead to the activation of SREBP. Specifically, a decrease in cell surface NKA levels has been shown to trigger the redistribution of a substantial portion of caveolin and cholesterol to the ER, subsequently promoting the activation of the SREBP pathway (40). Viewed from this angle, the proposed poisoning and depletion of functional raft domains by $PrP^{Sc}$ in prion diseases can be viewed as a trigger for cells to induce cholesterol synthesis. In agreement with this view, a profound upregulation of genes that mediate the biosynthesis of cholesterol or its transport were reported in prion disease paradigms (46, 47). Collectively, perturbations of raft domains with NKAs embedded in them will lead to $Ca^{2+}$ influx, localized increases in ATP, and the initiation of sterol biosynthesis (**Fig 1**).

## 2. Induction of ER stress response leading to UPR

With the perturbation of plasma membrane-resident raft domains, including the NKAs embedded in them, increasing over time—as we propose to be the case in a subset of brain cells in prion diseases—all the above rescue attempts will not achieve the objective to restore the essential electrochemical gradient and functional raft domains. This will drive cells to produce more replacement NKAs and other essential raft proteins. Consequently, the ER will become overcrowded with partially folded or unfolded proteins causing ER stress (48, 49). Cells are well-known to respond to this type of stress by inducing the unfolded protein response (UPR), a complex and ancient molecular program (50, 51). The UPR is often described to involve the highly coordinated activation of three response arms.

One arm of this response involves inositol-requiring protein 1α (IRE1α) and X-box binding protein 1 (XBP1). Specifically, ER stress induces the release of the 78-kDa glucose-regulated protein (GRP78), also known as BIP (52), from IRE1α. This release triggers IRE1α oligomerization and autophosphorylation, activating its endoribonuclease activity. The resultant activity excises a 26-nucleotide intron from XBP1 mRNA, causing a frameshift that leads to the translation of a 40 kDa mature XBP1 protein. The latter acts as a transcription factor, promoting the expression of genes that help cells cope better with ER stress, including the disulfide isomerase Erp57. Activation of the IRE1α-XBP1 pathway has been shown to increase the expression of $PrP^{C}$ (53). If the cell successfully generates enough folding capacity to resolve the ER stress, IRE1α can be inactivated. Although the precise mechanisms remain unclear, Erp57 has been reported to lower BIP levels, thereby decreasing IRE1α activation, which could be seen to reduce both $PrP^{C}$ and $PrP^{Sc}$ levels (54).

Another arm within the UPR is known to shut down general protein synthesis through the PKR-like ER kinase (PERK)-mediated phosphorylation of the eukaryotic translation initiation factor 2α (eIF2α). One study established a link between the inhibition of NKAs and the activation of this arm (55). Specifically, the inhibition of NKA was shown to stabilize a protein known as general control nonderepressible kinase 2 (GCN2), one of the kinases responsible for eIF2α phosphorylation. In other words, available evidence suggests that this arm of the UPR will be potentiated when the pump is inhibited. The translational repression afforded by eIF2α phosphorylation might be expected to be beneficial in prion diseases if it deprives cells of the $PrP^{C}$ substrate for conversion into $PrP^{Sc}$. Consistent with this expectation, an inhibitor of an eIF2α phosphatase has been shown to reduce $PrP^{C}$ expression and conversion (56). Regrettably, the translational repression also affects many other proteins needed for neuronal survival. As such, it has been repeatedly observed that eIF2α phosphorylation contributes to cognitive decline in prion disease, possibly because it reduces the production of essential synaptic building blocks. Although not all studies have observed translational repression in prion-infected paradigms (57), available *in vivo* evidence strongly favors the interpretation that stalled protein expression contributes to synaptic decay (58). Consistent with this view, reversing translational suppression genetically or through pharmacological inhibition of this arm of the UPR rescued synaptic deficits and memory deficits (58-61). Yet, these studies also established that the restoration of protein expression had a surprisingly small impact on prion-disease survival, a sign that these interventions did not address the underlying challenge that evolution has adapted to by slowing protein synthesis.

The reduction in overall protein synthesis afforded by eIF2α phosphorylation can be partially thwarted by the activation of a side branch of this adaptive UPR arm, which induces the expression of the activating transcription factor 4 (ATF4) and the CCAAT-enhancer-binding protein homologous protein (CHOP) (51, 58, 62-64). There are still many gaps in the understanding of the genes controlled by these transcription factors, yet available evidence suggests that they contribute to cell fate decisions that determine whether a cell responding to ER stress is going to activate programmed cell death or turn to selective protein translation for survival. If the latter outcome is favored, CHOP and ATF4 can restart the expression of selected genes responsible for the synthesis of amino acids and translation after they have been temporarily silenced by eIF2α phosphorylation (51, 65). This outcome may give a cell whose plasma-resident pumps are poisoned a fighting chance to replace the pump alongside its raft environment. Accordingly, a profound upregulation of CHOP was reported when NKAs were blocked through CGs (55, 66).

There are some inconsistencies in the literature regarding whether prion disease-afflicted cells induce CHOP. Transgenic mice that overexpress PrP D178N, although showing ER swelling (67), did not exhibit signs of UPR induction and, consequently, also failed to show increased levels of CHOP (68). However, a direct correlation between $PrP^{Sc}$ burden and CHOP levels was consistently observed in several other paradigms, including prion-infected neuroblastoma cells (54) and prion-infected mice (58). Moreover, the poisoning and stalling of raft domains within the ER through the transgenic expression of a construct comprising the flexible N-terminal tail of the prion protein fused to a GPI-anchor was reported to mimic ER-stress encountered in prion diseases that is characterized by phosphorylation of eIF2α and CHOP activation (69).

A subset of familial prion disease cases are caused by the respective prion protein mutants acquiring an unexpected membrane topology. Specifically, in these cases the prion protein has a propensity to get inserted as a transmembrane protein with its C-terminus facing the cytosol (${}^{Ctm}PrP$). When this ${}^{Ctm}PrP$ is encountered, the ensuing prion disease does not exhibit the conventional build-up of $PrP^{Sc}$ but shares with other forms of the disease a cellular demise characterized by spongiform degeneration. Any attempt to explain the molecular mechanisms underlying cell death in prion diseases should, therefore, account for the similarity in downstream cell death characteristics. To reconcile these events, one can refer to data showing that $PrP^{C}$ enhances the activity of NKAs (5). It can be speculated that ${}^{Ctm}PrP$—on account of no longer interacting normally with the pump and raft domains—initiates a cellular response that mimics the effects of latent pump inhibition. Alternatively, ${}^{Ctm}PrP$ may cause a breakdown of the electrochemical gradient through the formation of ion leak channels (70, 71), or by activating the UPR in a manner that resembles the molecular UPR fingerprint induced by $PrP^{Sc}$. Consistent with either scenario, increased CHOP expression was observed in cells engineered to express mutants that promote $^{Ctm}$PrP topology (72, 73).

A third arm of the adaptive UPR centered on the ATF6 transcription factor promotes—alongside IREα—the biogenesis of ER and Golgi compartments (74). Although less frequently monitored in published reports than CHOP or ATF4, studies which have explored ATF6 activation found it to be beneficial in relevant ER protein misfolding paradigms. When paired with the CHOP- and ATF4-mediated induction of selective protein synthesis, the promotion of ER biogenesis by ATF6 can be expected to reduce ER stress by increasing available folding space. Consistent with this anticipation, the activation of these proteins through overexpression, active mutants, or pharmacological agents was observed to reduce PrP aggregation in separate *in vitro* PrP misfolding paradigms (74, 75).

The UPR can be activated by a large variety of perturbations in protein homeostasis, which do not also cause massive PNS dilation. This indicates that additional conditions must be met before this phenotype develops. Although the relative balance of activation of the three branches of the UPR must be crucial, it is unlikely sufficient by itself for also generating the turgor within the PNS.

## 3. Massive osmotic swellings of ER and PNS

To gain purchase on the conditions that need to be met for PNS dilations to manifest, we will review in this section a subset of paradigms whose phenotypic hallmark are massive intracellular swellings of the ER and PNS. Because the PNS represents the largest continuous ER cisterna in most cells, profound ER swelling may most readily be observed as a PNS dilation. A molecular signature of these conditions seems to be the activation of an adaptive UPR, characterized by increases in CHOP levels—a prerequisite for sustained selective protein expression under ER stress—paired with $Ca^{2+}$ influx into the cell. With a view of identifying common themes, we found it helpful to sort these paradigms into five categories based on shared features amongst them (**Table 1** and **Figure 1**) (additional categories may be found when considering other paraptosis paradigms that have been published and reviewed (76)):

*Category I*: Insults affecting the pumping activity of NKAs. This category includes a) poisoning of certain cell types (for example, glia) with CGs (17, 77-79), b) hypoxic-ischemia, followed by ATP depletion (80), and c) exposure of cells to Tat-Beclin1—a 31 amino acid peptide generated by the artificial fusion of the HIV Tat peptide and a fragment of the autophagy inducing protein Beclin 1. Tat-Beclin1 has recently been proposed to exert its effect through interaction with NKAs (80-82).

*Category II*: Exposure of cells to selected toxins that broadly compromise the protein folding capacity of the ER or the shuttling of proteins out of the ER. In this category belong the poisoning of cells with: a) tunicamycin, a fungal toxin that blocks the first step in the synthesis of N-glycans (83, 84), b) ophiobolin-A (OphA), a compound that has been shown to form covalent thiol adducts in the ER, thereby broadly perturbing disulfide formation (85, 86), and (c) cyclosporine A (CsA), a cyclic peptide of 11 amino acids best known for its immunosuppressant uses that blocks the activity of peptidyl-prolyl isomerases of the cyclophilin family (87).

*Category III*: Overexpression of multi-spanning transmembrane proteins. Given the vastness of the protein overexpression literature, it is noteworthy that only a small number of overexpression paradigms have been reported to give rise to massive ER dilations, including a) truncated expression constructs of the lamin B receptor (LBR), a protein whose dual functionality comprises a role in anchoring lamin B to the inner nuclear membrane and an enzymatic function as a sterol reductase that catalyzes the reduction of the C14 unsaturated bond of lanosterol, b) 7-dehydrocholesterol reductase (DHCR7), the reductase responsible for the final step in the biosynthesis of cholesterol, namely the reduction of the C7-C8 double bond of 7-dehydrocholesterol, or c) transmembrane 7 superfamily member 2 (TM7SF2), the canonical reductase within the cholesterol biosynthesis pathway that, like the lamin B receptor, reduces the C14-unsaturated bond of lanosterol (88). In the course of our work, we showed that d) NKA α subunits and other closely related P2c-type ATPases, as well as (e) transient receptor potential vanilloid channel 2 (TRPV2) also belong in this category (24).

*Category IV:*: Other types of insults, including certain types of aberrant signaling. In this category belong: a) the expression of a C-terminal intracellular domain of the insulin-like growth factor I receptor (IGFIR-IC) (89), b) addition of tumor necrosis factor (TNF), or c) exposure of cells to polyinosinic:polycytidylic (90).

*Category V:* We place in this category the pore forming toxin (PFT) protein, aerolysin, expressed by the bacterium *Aeromonas hydrophila*.

At cursory inspection, it is difficult to see a common thread in these insults that may indicate why each of them leads to massive PNS dilations. This is particularly true for the Category IV insults. One may be tempted to conclude that a highly dense ER lumen, which may be caused by Category I, II, and III insults, is itself osmotically active, thereby driving water influx into the ER.

Category II compounds would seem to corroborate this notion because they share a profound influence on the ability of ER-directed proteins to progress in their passage of the secretory pathway past the ER. Thus tunicamycin, OphA and CsA have all been shown to cause stalling of the secretory pathway that would be expected to also affect nascent NKAs and other essential raft proteins that cells may try to replace. That this is not just a theoretical scenario was documented for CsA. Specifically, it was observed that CsA inhibits NKAs (91), not through a CG-like inhibition mechanism (92) but by preventing an interaction between the β1 pump subunit and cyclophilin B that was shown to be critical for maturation and trafficking of NKAs through the secretory pathway (93). In other words, CsA addition to cells would by itself be expected to trap nascent NKAs in the ER.

In our view, a lack of electron-dense structures in the dilated PNS cisternae contradicts the idea that protein densities themselves cause a sufficiently large osmotic imbalance to drive the massive PNS dilation phenotype. Yet, this argument does not diminish our view that these PNS dilations result from internal turgor, which frequently can be seen in paraptosis paradigms to compress the nucleus on the side adjacent to the ER. Considering that the osmotically active entity ought to be translucent, it seems most likely that it is caused by an ion imbalance across the ER membrane.

The most informative subgroup for understanding the mechanism underlying PNS dilation may be the Category III insults. A cursory look at this category begs the question why only the overexpression of certain proteins cause the PNS dilation? A common denominator of Category III protein overexpression paradigms seems to be that they are most often based on multi-spanning transmembrane proteins that contribute to sterol biosynthesis or act as ion channels, i.e., P2c-type ATPases and TRPV2 (24). Interestingly, these ion channels do not only contribute to the ion homeostasis across cellular membranes, but they share with sterol biosynthesis proteins their embeddedness in cholesterol-rich raft domains. In fact, a close functional relationship between NKAs was proposed to have been the main driving force for the evolution of cholesterol in metazoans (94). Validating a specific adaptation of NKAs to cholesterol are data from a yeast strain (*P. pastoris*) that was engineered to synthesize cholesterol, instead of the similar ergosterol that it normally assembles. This manipulation boosted NKA levels three-fold (95). Moreover, the treatment of cells with CGs was observed to also upregulate proteins responsible for cholesterol biosynthesis (95).

In search of an explanation for the turgor that drives PNS dilations we considered that the uneven antiport of two $K^+$ ions for every three $Na^+$ ions that characterizes each ATP-dependent NKA pumping cycle might gradually cause an ion imbalance across the ER membrane. This idea was refuted when we observed that the catalytic activity of NKA α subunits is not required for the PNS dilation to occur, because the overexpression of catalytically dead NKA α subunits could invoke the PNS dilation phenotype just as well as the overexpression of wild-type NKA α subunits (24).

A possible breakthrough in this pursuit provided data that we collected with the transient receptor potential vanilloid type 2 (TRPV2) channel. Overexpression of TRPV2 not only triggered the PNS dilation but pharmacological studies suggested that its $Ca^{2+}$ leak channel activity may directly contribute to the turgor formation (24). We had zeroed in on TRPV2 because this protein is known to act as a multi-modal ER membrane-resident $Ca^{2+}$ leak channel that combined several pertinent features: 1) TRPV2, formerly known as GRC, had been identified as an insulin-like growth factor-1 (IGF1)-responsive ER channel, which was observed to translocate to the plasma membrane in cells exposed to IGF1 (96). This was notable because we were aware that the Category IV group of insults leading to paraptosis included the expression of a C-terminal intracellular domain of the IGF1 receptor (89). In fact, this was the first report of this phenotype, leading the authors to propose the term paraptosis to describe it. 2) TRPV2 is known to be responsive to a range of toxins, including cannabidiol (97, 98), which had been shown to induce ER swelling (99). 3) It had been reported that overexpression of a TRPV2 paralog, TRPV1, can induce a phenotype that the authors characterized as paraptosis—regrettably, images included in this prior report did not allow us to assess if it has similarity to our phenotype, which is why we it did not pass our criteria for inclusion amongst the Category III overexpression paradigms (100). 4) A subset of TRPV channels were reported to function as ER-based mechano-sensors, making them fitting potential contributors to a phenotype characterized by osmotic swelling. 5) TRPV2 can also be activated by certain lipids, whose balance might be altered when members of the cholesterol synthesis pathway are overexpressed, as is likely the case in the DHCR7 and TMSF7 overexpression paradigm (88, 101).

Our subsequent experimental work, which largely made use of the OphA exposure paradigm (Category II) and the overexpression paradigms for NKA α and TRPV2, established that TRPV2 can temporarily translocate from the ER to the plasma membrane. We showed that PNS dilations can be antagonized by chelating $Ca^{2+}$, or by hindering the temporary translocation of TRPV2 to the plasma membrane, either pharmacologically through the addition of the TRP inhibitor tranilast or by expressing an N-terminal TRPV2 truncation construct. Moreover, we established a dual response of cells to SET2, a specific antagonist of TRPV2. Specifically, we demonstrated that SET2 can independently induce paraptosis, likely by inhibiting the natural ER-based $Ca^{2+}$ leak activity of TRPV2. However, SET2 also appears to reverse the OphA-induced paraptosis when introduced later and at a lower concentration. This effect may occur by blocking the transient $Ca^{2+}$ influx into the cell, which TRPV2 seems to drive after its translocation to the plasma membrane (**Fig 1**).

It seems to us that each of the five insult categories induce an adaptive UPR. This UPR is characterized by a sustained upregulation of CHOP driving selective protein translation and an aberrant accumulation of raft resident transmembrane proteins within the ER and PNS. Because $Ca^{2+}$ is a crucial co-factor for chaperone-mediated folding in the ER, the cell may have evolved to react to these insults by increasing its ER $Ca^{2+}$ levels. We have shown that in the OphA paradigm, cells achieve this increase in ER $Ca^{2+}$ levels by temporarily translocating TRPV2 to the plasma membrane, thereby 1) reducing its ER-based $Ca^{2+}$ leak activity in exchange for 2) boosting the influx of $Ca^{2+}$ into the cell. The osmotically active $Ca^{2+}$ counteracts the crowding in the ER and PNS by increasing the volume of these subcellular compartments.

We consider it likely that the PNS dilation phenotype can be induced in more than one way, so long as perturbations cause an influx of $Ca^{2+}$ into cells, paired with an ER stress that induces upregulation of proteins involved in sterol biosynthesis. This interpretation is validated by the Category V phenotype of ER vacuolation induced by the toxin aerolysin. Proaerolysin binds to GPI-anchored proteins on the plasma membrane. This binding is crucial for proaerolysin to unleash downstream toxicity because removal of GPI anchors by PI-PLC precludes the aerolysin-induced ER vacuolation (102). Once bound, proaerolysin is converted by endoproteolysis into aerolysin, which then forms a heptameric pore in the plasma membrane that permits $K^+$ efflux and $Ca^{2+}$ influx. This increase in permeability of the plasma membrane to small cations has been shown to be required for ER vacuolation. This was demonstrated by the failure to induce ER vacuolation of a mutant proaerolysin that can bind GPI anchors without altering ion permeabilities (102). Intriguingly, generic ionophores that increase $Ca^{2+}$ entry (e.g., ionomycin, A23187) or some other pore-forming toxins (e.g., streptolysin O) also do not induce ER vacuolation, suggesting that vacuolation requires more than just disturbances to ion balances (102). We propose that a crucial additional component for the PNS dilation to manifest is the concomitant activation of SREBP. As we had observed in our experiments based on the OphA paraptosis paradigm, the addition of $Ca^{2+}$ chelators to the extracellular space prevented the aerolysin-dependent ER vacuolation (103). However, the authors also documented that these chelators did not inhibit the activation of the aerolysin-induced activation of SREBP. We propose that the lack of $Ca^{2+}$ under these conditions may trigger a related phenomenon of ER hyperproliferation.

## 4. Induction of sterol biosynthesis and ER hypertrophy

To restore functional raft domains, the cell must not only replace the embedded proteins but also the cholesterol and sphingolipids that are enriched in these specialized membrane domains. For instance, blocking the synthesis of cholesterol by the addition of the HMG-CoA-reductase inhibitor lovastatin (also known as mevinolin) has been shown to trap PrP in the ER (104). If the regular biosynthesis of raft domains does not restore cell surface levels required for cell survival, the cell can escalate their production by rapidly producing additional ER compartments, a phenomenon known as ER hypertrophy. This phenomenon was first observed to exist in cells exposed to certain toxins, including phenobarbital (105). In the latter context, it was proposed to serve the purpose to enhance the capacity of cells to detoxify themselves through the massive production of ER-resident proteins that can disarm the respective toxin. In subsequent work, it was shown that this program is associated with CHOP activation. Moreover, the mere overexpression of certain ER-resident transmembrane proteins is sufficient to induce an ER hyperproliferation phenotype. Elegant studies were undertaken with a Chinese hamster ovary (CHO) cell line (UT-1), obtained by stepwise adaptation to growth in the presence of compactin, a fungal metabolite known to inhibit HMG-CoA reductase (HMGR). HMGR is the rate limiting reductase within the cholesterol biosynthesis pathway and its levels in UT-1 cells exceeded 500-fold normal levels, causing the ER in these cells to be packed in crystalloid arrays (106). Subsequently, it was proposed that upon addition of compactin to UT-1 cells, the outer perinuclear membrane initially buds and detaches from the inner membrane, and subsequently morphs into tightly packed juxtanuclear ER stacks (107), later referred to as karmellae (108).

Since then, the overexpression of several transmembrane proteins with known roles in the biosynthesis of cholesterol have been shown to induce ER hypertrophy (109). It is now understood that karmellae can appear surprisingly fast (within minutes) and can be dynamic. Similar ER stacks can also be seen adjacent to the plasma membrane or within the cytoplasm. The shape and fine structure of these ER proliferations is influenced by the nature of the protein that is overexpressed to induce them and by the angle from which they are viewed. Frequently observed are parallel stacks, whorls and near crystalline arrangements (107, 110). Moreover, the molecular biology that governs ER hyperproliferation is present in a large variety of cell types (111).

To date, more than a dozen manipulations are known to give rise to hypertrophied ER, also known as organized smooth ER (OSER) (109, 112-114). Mutagenesis studies on hybrid artificial proteins have led to the conclusion that there is not a common sequence shared amongst transmembrane proteins that can induce OSER structures. Yet, brute force overexpression of a given ER-resident membrane protein alone does not guarantee OSER structure formation. Rather, for a highly expressed protein to induce this phenotype, it needs to be able to engage in low-affinity homotypic protein interactions *in trans*, i.e., from one ER cisternae to another across a cytosolic gap (113, 115).

Remarkably, although several artificial proteins have been engineered that can induce OSER structures, natural proteins that give rise to this phenomenon share a conspicuous relationship with cholesterol. Thus, they are either predominantly the enzymes required for the biosynthesis of cholesterol, are known to induce cholesterol-enriched raft domains, reside within them, or influence their transport to the cell surface. With this information in mind, we visited early data on cholesterol synthesis inhibitors, which were already available in the 1960, including the DHCR7 and DHCR24 inhibitors AY-9944 and triparanol, respectively, and found several reports that captured in ultrastructural analyses morphologically indistinguishable lamellar structures that we can now interpret to most likely also represent ER hypertrophy (116, 117).

Given the many similar conditions that can invoke massive ER dilation and ER hypertrophy, with the latter mainly lacking the apparent requirement for $Ca^{2+}$ influx, we considered that these phenotypes may be related events. Indeed, we found corroborating evidence for this notion in reports that experimentally induced these phenotypes through overexpression of sterol reductases. For instance, the overexpression of LBR constructs that differ in one amino acid at the C-terminus have been shown to give rise in separate cells to ER dilation or ER hypertrophy (88, 118) (**Supplemental Fig 1A**). Moreover, evidence of OSER-like hypertrophy (although not explicitly highlighted by the authors) can be detected in cells that were chronically exposed to CGs, for example, Fig. 8 in (119).

## 5. Evidence for PNS dilation and ER hypertrophy, in the neuropathological record of prion diseases.

Our recent work documented direct *in vivo* evidence of PNS dilation in prion-infected mice (24). Specifically, the experiment made use of mice, which were infected with RML prions at 60 days of age and transduced with recombinant adeno-associated virus (rAAV) vectors at 90 days post-inoculation (dpi). The payload of the rAAV vectors coded for an enhanced green fluorescent protein (EGFP), which we directed to the ER and PNS by fusing an ER signal peptide to its N-terminus and a 'KDEL' ER retention motif to its C-terminus (spEGFP$^{KDEL}$). Using fluorescence microscopy of cryo-sectioned brain sections, we showed that in prion-infected—but not in non-infected control mice—a small number of vacuoles in immediate proximity to nuclei were filled with spEGFP$^{KDEL}$, making them recognizable as PNS dilations (**Fig 2**).

In parallel to undertaking these studies, we reviewed and compared pathological records of prion diseases against known subcellular vacuolation phenomena, including reports depicting swelling of endosomes and lysosomes, mitochondria, multi-vesicular bodies, autophagosomes, the Golgi or the ER. In our survey of the pertinent literature, we initially focused on images of brain sections from experimental prion disease studies that can give rise to the best-preserved specimens due to shorter postmortem times than those typically achieved with human specimens. There, we found a plethora of histopathological images depicting what the authors referred to as 'spongiform degeneration' resembling the vacuolation we could find in images of cells that were experimentally induced and verified to exhibit profound PNS dilations (**Fig 2A,B**). In fact, one study of marmosets injected with prions led its authors to describe the vacuoles as dilations of ER cisternae (120). We also detected abundant evidence for its occurrence in the human pathological record of prion disease (**Fig 2C**) or natural and experimentally induced animal prion diseases (**Fig 2D,E**).

Once we had formulated the hypothesis that PNS dilations and ER hypertrophy may represent related phenomena, we anticipated that evidence for ER hypertrophy might also feature in the neuropathological record of prion diseases. Indeed, we found evidence to this effect. For instance, incidences of membrane hypertrophy were emphasized by a subset of authors not only in early but also in recent prion disease reports (121, 122) (**Fig 3**). The images provided in the respective reports leave no doubt as to them representing either OSER structures or hyperproliferated rough ER.

How significant are PNS dilations and ER hypertrophy for the late-stage pathology of prion diseases? Another neuropathological feature that has been widely agreed upon to exist in prion-infected brain sections but not in controls may provide an additional hint regarding the significance of these ER perturbations for the neuropathology of prion diseases.

## 6. Association of PNS dilation and ER hypertrophy with paracrystalline aggregates of inactive ribosome

At the time when the prion hypothesis had not yet been formulated, several leading neuropathologists independently published negative stain electron microscopy images of prion-infected brain sections, which showed concentrations of 'dense particles' and 'rod-like' structures of a similar diameter (23-37 nm) that sometimes were seen to acquire vermicular or paracrystalline appearance (121, 123-128) (**Fig 4, panels A,C, and E**). The authors of these articles agreed that this ultrastructural feature was not observed in non-infected brains. Subsequent work by others corroborated this conclusion (129, 130) and emphasized their 'virus-like' characteristics (131). To our knowledge, the identity of the molecular entity forming these structures has not been resolved to date.

When exploring the PNS dilation literature, we noticed electron micrographs of liver cells in hypothermic chickens. In addition to distended PNS cisternae, these images revealed paracrystalline arrays like those present in the prion disease literature (132) (**Supplemental Fig 1B**). The authors had initiated the study to learn more about the phenomenon of ribosome crystallization. A deeper dive into the small literature describing this phenomenon then revealed that ribosome crystals were initially observed in 1966 in the embryonic lens of 10-day-old chick embryos, which had been grown under hypothermic conditions (133) (**Fig 4, panels B and D**). In this milestone paper, ribosomes formed a lattice characterized by a repeating unit composed of four ribosomes that come together in a square array.

The biochemical characterization of purified ribosome crystals from hypothermic chick embryos revealed subsequently that these contained the same rRNA as normal ribosomes but were shown to lack tRNAs and mRNAs (134). Moreover, it was demonstrated that the slow cooling deployed had permitted the elongation and termination of nascent polypeptides but prevented the initiation of new chains. It is now understood that ribosomes, which are not loaded with mRNA or nascent proteins, exhibit under hypothermic conditions a tendency to aggregate into tetramers after approximately two hours. However, if protein synthesis is inhibited before the slow hypothermic cooling is initiated, then the tetrameric formation may be replaced with structurally less ordered arrangements (135). Since then, paracrystalline ribosome assemblies have been observed in other conditions, including hibernation (136). Consequently, evolution may have equipped cells with an ability to store ribosomes in a paracrystalline array when not needed.

In the disease context reviewed here, brain cells that react to the poisoning of raft domains with an attempt to replace them will not only be induced through the adaptive UPR to undergo ER dilation or ER hypertrophy through selective translation and ER biogenesis but can initially also be expected to ramp up the biosynthesis of ribosomes. As the cellular health deteriorates supply lines for keeping up the fight will eventually come to a halt. This outcome may be reached sooner if the outer and inner nuclear membranes detach from each other during the distention of the PNS, thereby temporarily or permanently disrupting the shuttling of mRNAs and tRNAs to the cytoplasm. Under these circumstances, it may not take long before ribosomes become inactive and unoccupied with mRNAs, tRNAs, and nascent proteins, thereby favoring their aggregation into paracrystalline assemblies. If this course of events is initiated at normal body temperature in late-stage disease, the crystal lattice can be expected to be less organized and may show little evidence of the tetrameric unit found in aggregates that form under hypothermic conditions.

The side-by-side comparison of late-stage prion disease brain sections with published images from paracrystalline ribosome aggregate studies reveals several facets supporting this model: 1) The sizes of the paracrystalline assemblies that form in both instances are consistent. 2) The negative staining characteristics of the aggregates are comparable. 3) In both scenarios, the aggregating units can be observed to assemble strings of similar length and sometimes can be seen to form rounded structures exhibiting similar curvatures (**Fig 4, panels G and H**).

The idea that paracrystalline structures observed in prion diseases may comprise ribosomes was considered before but was dismissed because the particles in them were described to exhibit 'sharper outlines' (131). We suggest that this astute observation of differences in the appearance to active ribosomes may merely reflect the absence of tRNA, mRNA, and nascent protein chains in the inactive paracrystalline assemblies (134).

If the proposed relationship between PNS dilation and ER hypertrophy has merit, one may expect the literature to also provide incidences of ribosome crystallization being reported in cells with OSER structures. The latter is indeed the case—despite the small body of literature describing this phenotype. Specifically, the infection of cells with herpes viruses was documented to manifest ultra-structurally in the appearance of ribosome crystals and ER hypertrophy (137, 138) (**Supplemental Fig 1C**). The studies were again published prior to the designation of these highly organized subcellular compartments as OSER structures. However, the images presented in the respective reports leave little doubt regarding their identity.

Returning to the question of the significance of PNS dilations for the pathobiology of late-stage prion disease, one may formulate the following indirect answer: If the proposed relationship between PNS dilations, ER hypertrophy, and paracrystalline inactive ribosomes holds merit, then the consistency with which the appearance of these structures was reported in prion-infected brain tissue would suggest the biology underlying their formation to be a distinguishing event in the late pathobiology of prion diseases. This event cannot be rare given that several groups documented it independently despite the small size of brain tissue areas that would have been sampled by electron microscopy (123-127, 139). For instance, in one study co-authored by the NIH prion research team at Bethesda, 9 out of 33 scrapie brains contained varicose tubules and the same number of mice contained round or oval bodies, which we now propose to represent ER hypertrophy. The study authors also pointed out that 14 out of 33 scrapie mice contained dense osmiophilic particles, which we interpret to represent inactive ribosome paracrystals (121). Remarkably, no other distinguishing features were identified in prion-infected brain sections by these authors and none of these structures were detected in control mice.

Why has it been so difficult to determine unequivocally which subcellular compartment gives rise to vacuoles that underlie the spongiform degeneration phenotype? In our view, certain characteristics of PNS dilations may have contributed to them not having been robustly identified as a possible source for these vacuoles: 1) Cells exhibiting PNS dilations may be relatively short-lived. 2) We have shown that conventional methods of specimen preservation destroy their structural integrity (24). Specifically, when we directed spEGFP$^{KDEL}$ to these cisternae, even fixation with 0.5% formaldehyde for 30 minutes precluded detection of the green fluorescence signal within their lumen. We observed that the fragility of PNS dilations correlated with their size, because smaller ER vesicles and PNS dilations retained the spEGFP$^{KDEL}$ signal even when fixed more aggressively. Consequently, to date, we have only been able to label smaller PNS dilations in prion-infected mice by directing spEGFP$^{KDEL}$ into their lumen (**Fig 2K**).

We propose that additional complexity stood in the way of recognizing a role of ER dilation and ER hypertrophy for the pathogenesis of prion diseases. Time-lapse videos, which we recorded of cells that were chronically exposed to OphA document that the cells eventually burst. Intriguingly, this process appears to follow two steps. Initially, the cells disintegrated into smaller membrane-surrounded vesicles. Some of these randomly formed lipid vesicles stuck together creating a 'soap bubble effect' (121) and surrounded portions of the prior cellular content.

At a later stage, the 'soap bubbles' often fuse to larger membrane-surrounded 'containers' of the prior cellular content. Finally, as the barrier function of the lipid membrane that encloses these containers breaks down, they can be seen to inflate—presumably due to the osmotic differential across the membrane. When we directed spEGFP$^{KDEL}$ to the ER and PNS, the green fluorescence-filled dilations could sometimes be observed to survive the initial burst and release their content only during the late-stage dilation. This sequence of events was recorded with a human osteosarcoma (U2OS) cell line by live cell microscopy (not shown). If it translates to differentiated neurons remains to be determined.

Regardless of how cell death manifests in neurons, once they succumb to injury and the structural integrity of cellular compartments collapses, the release of proteolytic enzymes and nucleases is inevitable. This will cause the degradation of the enzymes' respective substrates. The tightly packed assembly state of paracrystalline inactive ribosomes and OSER structures can be expected to confer resistance to degradation, thereby favoring their subsequent detection. Moreover, as the cells burst, they will get severed from their processes which may be filled with these same structures because the cells may have previously undertaken a futile attempt to restore synaptic functionality. The detachment of these processes will deprive ribosomes and ER that exist within them of tRNAs and mRNAs supplies, thereby recreating an environment favoring the formation of paracrystalline aggregates of inactive ribosomes. As the cellular processes deteriorate, the osmotic pressure differential across their plasma membrane will cause them to swell, leading to a loss of typical morphology. As a result, post-synaptic spines will appear larger and more rounded than typical.

Consistent with this course of events, the paracrystalline ‘virus-like’ particles, described in the neuropathological record of prion diseases, were often detected in rounded vesicles that also included a synaptic cleft, allowing them to be classified as enlarged post-synaptic densities. If key steps of the cellular decay that we observed in U2OS cells translate to neurons, ER and PNS dilation-dependent spongiform degeneration can be expected to give rise to two types of liquid-filled spaces, the ER and PNS-derived vesicles and artificial bubbles that either originate from the initial bursting of the plasma membrane or the subsequent dilation of the lipid-bound container. In line with the chronology inherent to this model, we observed that ER and PNS dilations are best captured in monkey or mouse brain tissue in experimental prion disease paradigms (presumably because the animals were sacrificed before the disease took them), whereas structures resembling plasma membrane-derived liquid-filled spaces, membrane fragments, and inflated post-synaptic densities appear to contribute a larger share of reported observations in human post-mortem brain sections.

We would be remiss if we did not mention the alternative explanation for the identity of vacuoles in prion diseases that has dominated the spongiform degeneration literature, namely the suggestion that they develop from endolysosomal compartments. This possibility has recently received additional attention when it was reported that levels of 1-phosphatidylinositol-3-phosphate 5-kinase (Pikfyve) were greatly diminished in several mouse prion disease paradigms, including in prion-infected mice (140). Intriguingly, we have observed that PIKFYVE levels are also significantly reduced in human cells that evidently exhibit dilated ER and PNS cisternae because we had treated them with OphA and had directed spEGFP$^{KDEL}$ to their lumen (24). For now, we ascribe the lowering of steady-state PIKFYVE levels in OphA-treated cells to stalled protein expression driven by UPR activation. Critically, these results raise concerns about deductive reasoning by which reductions in PIKFYVE levels automatically point to endolysosomal origins of vacuoles. To settle the possible contribution of PIKFYVE to spongiform degeneration, it will be necessary to document that the vacuoles observed in prion-infected mice can be filled with an endolysosomal marker.

Are there other characteristics of PNS dilations that can be explored to determine if spongiform degeneration in prion diseases are of PNS or endolysosomal origins? We considered three characteristics of PNS dilations that should be less often encountered if spongiform degeneration was of endolysosomal origins, namely, their 1) inherent proximity to the nucleus, 2) the observation that the turgor in them tends to deform the nucleus such that the side of the nucleus abutting them is concave, and 3) a correlation between the size of the PNS dilation and the compaction of the adjacent nucleus, an observation we have repeatedly made in PNS dilation paradigms (please see top and bottom panels of **Fig 2I** for an example). Consequently, we compared H&E stains of spongiform degeneration caused by experimental prion disease in mice with vacuoles observed in mice deficient for the Pikfyve interactor polyphosphoinositide phosphatase (Fig4) (**Fig 5**). Because these data were generated by independent groups using slightly different histopathological procedures and brain areas, the published images could not be perfectly matched. In both types of histopathological samples, one can find examples of large vacuoles that are not adjacent to nuclei. This is not surprising because the most common thickness of brain sections for H&E staining is 5 μm yet PNS dilations can easily exceed 20 μm in diameter, which can result in images of PNS dilations not showing the nuclei with which they are associated. Regrettably, the number of suitable images available that depict endolysosomal vacuolation paradigms is relatively small, precluding a meaningful statistical comparison. These caveats aside, here is what we observed: 1) vacuoles in prion disease tissues are more often seen adjacent to nuclei than endolysosomal vacuoles generated by Fig4-deficiency, 2) prion disease vacuoles more often compress nuclei adjacent to them than endolysosomal vacuoles, and 3) prion disease vacuoles are more frequently seen next to compacted nuclei than endolysosomal vacuoles. Taken together, these observations favor the interpretation that the spongiform degeneration in these prion disease paradigms was predominantly caused by ER and PNS dilation rather than endolysosomal vacuolation.

## Conclusions

Our initial impetus had been to understand if a phenotype of PNS dilations contributes to spongiform degeneration, a hallmark of prion diseases. During this exercise, it became apparent that the PNS dilation phenotype manifests when cells are exposed to certain insults. We grouped these insults into five categories and propose that each of them can be expected to trigger:

1) The induction of an adaptive UPR in a way that emphasizes selective translation through CHOP/ATF4 activation.

2) Upregulation of ER biogenesis and sterol biosynthesis.

3) A cellular biology that increases cellular $Ca^{2+}$ influx through the plasma membrane.

We propose that this $Ca^{2+}$ influx, also observed by others in prion disease paradigms (141), may manifest in more than one way in the same or different paradigms. Likely contributors are the reverse mode activity of NCX, influx through the NMDA receptor, the formation of toxic non-selective ion channels at the plasma membrane, or temporary $Ca^{2+}$ influx through TRPV2 calcium leak channels (**Fig 6**).

We propose that prion diseases are accompanied by a gradual poisoning of raft domains, which can be expected to have a pronounced effect on the proteins within them, including NKAs. Because the vulnerability of brain cells to the accumulation of abnormal conformers of the prion protein versus acute CG poisoning is bound to be non-identical, the brain pattern of vacuolation induced by either insult may be overlapping but should not be expected to be identical. We reminded readers that the evolution of NKAs is intricately linked to the evolution of cholesterol (94). As a result, when cells upregulate NKA expression in response to poisoning, they also activate the sterol synthesis pathway. While we expect the impairment of NKAs to be a main driver of a rescue biology triggered late in the disease and deemed it helpful to review the biology of NKAs poisoning for model building, other raft-resident proteins can be expected to contribute to feedback mechanisms that prompt neurons to restore functional raft domains.

We also pointed to evidence suggesting that the ER and PNS dilation phenotype is related to the phenomenon known as ER hypertrophy, which can lead to OSER structures. The molecular underpinnings favoring one outcome over the other are currently unknown. We propose that contributing factors include the rate at which $Ca^{2+}$ accumulates in the ER and PNS, as well as the activation of CHOP and SREBP. Whereas the former may promote the buildup of turgor, the latter triggers selective translation, leading to ER biogenesis and sterol biosynthesis. If these cellular responses fail to counteract the insult—such as when toxic levels of $PrP^{Sc}$ persist—the swelling of the ER and PNS, or continued ER hypertrophy, will lead to cell death.

We propose that the dense particles described in the neuropathological record of prion diseases, which have been the topic of much speculation over the years due to characteristics that some authors considered virus-like, represent paracrystalline arrangements of inactive ribosomes. These structures may be particularly resistant to the uncontrolled breakdown of cellular content that is prone to occur once the cellular integrity is compromised.

The model builds on the work of leading prion research teams spanning more than 50 years of discovery, which have repeatedly documented these neuropathological features. The model is strengthened by the relationships between these features that can also be observed together in other paradigms. Formal validation of the model will be multi-faceted and may include some relatively straightforward experiments, e.g., the labeling of paracrystalline assemblies of dense particles in prion disease with nanogold-conjugated antibodies that recognize ribosomes. It also will need to be determined to which extent the ER and PNS dilation versus endolysosomal vacuolation contributes to spongiform degeneration in prion diseases. More interesting will be to clarify the relative significance of the neuropathological features discussed for neuronal death in prion diseases with a view to generate significant new ideas for slowing the disease. Once the molecular underpinnings that steer neurons toward ER and PNS dilation versus ER hypertrophy are better understood, it might be possible to direct cells to whichever of these alternative cellular responses is more beneficial for survival. Doing so may win valuable time, particularly, when paired with treatment modalities that lower $PrP^C$ levels or block its conversion.

**Terminology**

_Spongiform degeneration_: A pathological condition in the brain characterized by the formation of microscopic vacuoles (fluid-filled spaces) within neurons, leading to a sponge-like appearance of brain tissue, first described in 1898 in sheep afflicted with Scrapie disease (142). This degeneration is commonly associated with prion diseases and is a hallmark but neither an exclusive nor invariable feature of these disorders.

_Status spongiosus_: In late-stage prion disease, brains often present with widespread cell loss, a feature known as "status spongiosus." This feature is shared with other neurodegenerative disorders, particularly frontotemporal lobar degeneration (FTLD)-spectrum diseases, which can exhibit similar cell loss (143, 144). However, whereas status spongiosus in prion diseases may occur throughout the cerebral cortex, it is more typically restricted to the first two cortical layers in FTLD (143, 145, 146), referred to as superficial linear spongiosis (SLS). This type of cell loss is also seen in corticobasal degeneration (147), Pick's disease (148), argyrophilic grain disease (149), and dementia with Lewy bodies (150). Additionally, status spongiosis can result from vascular, metabolic, or hypoxic encephalopathies (151-153), brain edema (154) and artefacts relating to tissue processing (155), reviewed in (156, 157).

_Oncosis_: This term, derived from ónkos (largeness) and ónkosis (swelling) was introduced in 1910 by F. Recklinghausen (158, 159), and has typically been used to characterize the cell death that ensues following a collapse of the electrochemical gradient or the depletion of cellular ATP, with ischemia being the most common condition.

*Paraptosis*: Coined in 2000, this term was initially introduced to describe novel features of cell death observed in cells made to overexpress the intracellular domain of the IGF1 receptor, including 1) cytoplasmic vacuolation that predominantly affected the ER, 2) resistance to pan-caspase inhibitors, 3) yet dependence on caspase 9 (89). Subsequent publications, which adapted this term, reported insults to cells that led to varying degrees of ER dilation. However, only a subset of them showed pronounced PNS dilation or investigated the unique profile of caspase independence and caspase 9 reliance described in the original report.

*Necroptosis*: A cell death program, first described in 2005, which appears to have evolved in higher vertebrates as a defense against viral agents that evade the host cell's apoptosis machinery (160). Like apoptosis, this program can be triggered by the stimulation of cell surface receptors, including TNFR, but can also be induced by intracellular sensors, e.g., the detection of viral RNA (161). Critical components of this death pathway are well understood and include RIPK1, its paralog RIPK3, and mixed-lineage kinase domain-like protein (MLKL) (162). Although necroptotic death induces autophagy, it does not seem to rely on this cellular program (160). Rather than being a general feature of necroptosis, massive ER swelling may be the exception with this type of cell death.

*Autosis*: This term was proposed in 2013 to describe the Tat-Beclin 1-induced cell death (see Category I insults) that untypically combines ER dilation with signs of autophagy (80). It remains to be established whether autosis represents a specialized version of oncosis and whether the autophagic characteristics of this cell death manifest independently of the ER vacuolation, perhaps on account of Beclin 1 playing a separate role in the induction of autophagy.

## Funding

This project was supported by an operating grant of the Canadian Institutes for Health Research (CIHR) (grant number 202209PJT). GS gratefully acknowledges most generous support from the Krembil Foundation. HW's contributions to this study were enabled through funding from the National Institute of Allergy and Infectious Diseases (NIAID) of the National Institutes of Health (NIH) under award number RO1AI156037.

## Figure Legends

**Figure 1: Cellular homeostasis of raft domain perturbation and NKA inhibition.**

Cartoon summarizing insults that may give rise to a common PNS dilation phenotype. For illustration purposes, the panels omit many details, some of which are discussed in the text. Because cellular components and signaling events are shown in a generic cell, some proteins are depicted with generic names, e.g., connexin, calpain, when different cell types may express distinct paralogs or isoforms of these proteins. Finally, only a subset of events depicted will manifest in each cell type.

**Figure 2: Evidence of PNS dilation in neuropathological images of spongiform degeneration in prion disease versus paraptosis in cells and mice.**

(A) Spongiform degeneration in hematoxylin-eosin (H&E) stained brain section of experimental GSS in a marmoset. Image obtained from (120), Page 1985, and reproduced with permission from Oxford University Press. Insets '1', '2', '3' and 'C' in this panel and in Panels 'B', 'C', and 'F' designate examples of vacuolation that increase in size or are characterized by a complex phenotype of multiple vacuoles abutting a single nucleus, respectively. (B) H&E-stained brain section of spongiform degeneration in mouse-adapted RML prion disease. Note the similarity of vacuolation characteristics shown in insets '1' and 'C' to those depicted at ultrastructural resolution in the top and bottom image of experimental PNS dilations depicted in Panel 'I', indicated by the dotted lines connecting these subpanels. (C) Spongiform degeneration in human variant Creutzfeldt-Jakob disease depicted by H&E staining (https://commons.wikimedia.org/wiki/File:Variant_Creutzfeldt-Jakob_disease_(vCJD),_H%26E.jpg ID#10131). (D) Electron microscopy image of a condensed nucleus in a case of natural scrapie in sheep, showing disrupted nuclear envelope. Imaged obtained from (122), Page 80, and reproduced with permission from Springer Nature. Note that the appearance of condensed ('pyknotic') nuclei is also a common feature of apoptosis but these nuclei being surrounded by large PNS dilations, as seen here, is not. (E) Spongiform degeneration in hippocampal formation of transgenic mouse expressing the cervid *Prnp* gene infected with chronic wasting disease (CWD) prions, characterized by shrunken neuronal nuclei adjacent to foci of spongiform change. Image obtained from (163), Pages 11347, Figure 1B, and reproduced with permission from the American Society for Microbiology (https://journals.asm.org/journal/jvi). (F) Ouabain-poisoned primary neural cells dissected from *Atp1a1*$^{s/s}$ mouse embryos expressing a cardiac glycoside (CG)-sensitive NKA α1 subunit. Reproduced from (24) under CC-BY license. (G) Live-cell microscopy image of human U2OS cells transfected with an ATP1A2 overexpression construct whose translation was linked to the expression of an enhanced green fluorescent protein (EGFP) through a 2A peptide (P2A) sequence derived from the porcine teschovirus (P2A) known to induce ribosomal skipping during translation. The latter was directed to the ER with a signal peptide and retained in this compartment by a C-terminal KDEL sequence (spEGFP$^{KDEL}$). The panel shows an overlay of a differential interference contrast (DIC) image of a PNS dilation caused by the overexpression of the NKA α subunit which is filled with the green fluorescent signal emitted by spEGFP$^{KDEL}$, thereby identifying the vacuole as an ER cisterna. Reproduced from (24) under CC-BY license. (H) Alternative depiction of the PNS dilation induced by overexpression of the ATP1A2-P2A-spEGFP$^{KDEL}$ construct in formalin-fixed U2OS cells. In this image, the ER and PNS boundaries are detected by an anti-KDEL antibody whose binding is visualized through the red fluorescence emitted by a secondary Cy5-confugated antibody. Note that the fixation released the content of the PNS dilation. Reproduced from (24) under CC-BY license. (I) Electron microscopy images of early-stage (top) and late-stage (bottom) paraptosis in U2OS cells overexpressing a C-terminally truncated lamin B receptor (LBR) (1-533) fused to a yellow fluorescent protein (YFP). Note that the black boxes and labels shown in these panels are from the original publication. Images obtained from (88), Page 363, Figure 7A1 and B1, and reproduced with permission of the American Society for Cell Biology, conveyed through Copyright Clearance Center, Inc. (J) Hippocampal formation of a ouabain-poisoned *Atp1a1*$^{s/s}$ mouse brain. Note the similarities in the morphologies of vacuoles seen in Panel 'E' (24). (K) Direct fluorescence evidence of early-stage PNS dilations (marked with red arrowheads) in the brainstem of RML prion-infected *C57Bl6* mice transduced with a recombinant adeno-associated virus (rAAV) vector coding for spEGFP$^{KDEL}$. The blue signal in these images marks nuclei and was generated by Hoechst DNA staining (24). Except for Panels B and J, which represent previously unpublished data by the authors, all images in this figure are authorized reproductions from the published reports cited. To emphasize morphological features relevant to this review, a subset of the original images was enlarged, turned, or cropped. Where sizing bars were present in the original figures, approximate sizing bars were also inserted here. No other manipulations were undertaken.

**Figure 3: Evidence of ER hypertrophy in the neuropathological record of prion diseases.** (A) Cartoon of pleomorphic or varicose, lamellar, and whorl-shaped ER hypertrophy. (B) Electron micrograph of mice infected with the Chandler prion strain showing a density pattern consistent with pleomorphic ER hypertrophy. Image obtained from (121), Page 316, Figure 4, and reproduced here with permission conveyed through Copyright Clearance Center, Inc. (C) Pleomorphic 'anastomosing' ER hypertrophy induced by overexpression of an artificial construct composed of mitochondrial cytochrome b(5) fused to green fluorescent protein (GFP). Image obtained from (113), Page 261, Figure 5D inset, and reproduced here under CC BY-NC-SA 4.0 license. (D) Electron microscopy image of highly organized lamellar structure in Chandler prion-infected mice. Image obtained from (121), Page 318, Figure 5, and reproduced here with permission conveyed through Copyright Clearance Center, Inc. (E) Lamellar ER hypertrophy in filamentous fungus. The partial white arrow visible in the image was already present in the review article from which this image was obtained (164). The image was originally published by (165) and is reproduced here under CC-BY 4.0 license. (F) Oval-shaped structure in Chandler prion-infected mice. Image obtained from (121), Page 318, Figure 7, and reproduced here with permission conveyed through Copyright Clearance Center, Inc. (G) Whorl-shaped OSER structure in CV-1 cells that were transiently transfected with cytochrome b(5) expression construct. Image obtained from (113), Page 259, Figure 2a, and reproduced here under CC BY-NC-SA 4.0 license. All electron micrographs shown in this figure were reproduced from the published manuscripts cited. Some of these images were turned and cropped to emphasize similarities of morphologies between neuropathological features observed in prion disease and in ER hypertrophy.

**Figure 4: Striking similarities between the morphology of dense 'virus-like' particle aggregates in prion disease and paracrystalline aggregates of inactive ribosomes.**

(A) Ultrastructural analysis of brains of Chandler prion-infected mice depicting dense particles. Note that for this and other images in this figure the interpretations that were published by the original authors is shown in quotation marks. Image obtained from (121), Page 315, Figure 2, and reproduced here with permission conveyed through Copyright Clearance Center, Inc. (B) Ribosome crystals in mitotic cell from a chick embryo cooled for 6 hours in 5°C chamber. Image obtained from (166), Plate X, and reproduced here with permission from Elsevier. (C) Paracrystalline arrangement of dense particles in Chandler prion-infected mouse brain. Image obtained from (124), Page 209, Figure 4A, and reproduced here with permission from John Wiley and Sons. (D) Alternative arrangement of inactive ribosome crystals observed in same paradigm as in Panel B. Image obtained from (166), Page 162, Plate I, and reproduced here with permission from Elsevier. (E) Electron micrograph of dense particles in brain of 'Klenck' scrapie prion-strain infected 5-month-old mouse. Image obtained from (126) and reproduced here with permission of SAGE Publications. (F) Other arrangement of ribosome crystals in same paradigm as Panel B. Image obtained from (166), Plate X, and reproduced here with permission from Elsevier. (G) Vermicular arrangement of dense particles observed in Chandler prion-infected mouse brain. Image obtained from (124), Page 208, Figure 3, and reproduced here with permission from John Wiley and Sons. (H) Polysomes observed in follicular epithelial cell. Image obtained from (167), Page 122, Figure 5, and reproduced here with permission from John Wiley and Sons. When comparing Panels G and H, note the striking similarities in the parallel arrangements of polyribosomes forming curls and turns. The black arrows shown in Panels B, F and H were placed by the authors of the original publications. Some of these images were turned and cropped to emphasize similarities of morphologies between neuropathological features observed in prion disease and paracrystalline arrangements of inactive ribosomes.

**Figure 5: Prion disease vacuoles can be distinguished from endolysosomal vacuoles by their relationship to nuclei.**

(A) Algorithm for semi-quantitative evaluation of relationship between vacuoles and nuclei in mice afflicted with prion disease or exhibiting endolysosomal vacuolation. (B) Low level vacuolation in the cerebral cortex of a mouse that had been inoculated with 22L prions. Note how the vacuole indicated with the black arrowhead seems to compress the adjacent nucleus and causes its compaction, recognizable by its darker hematoxylin staining than surrounding nuclei. This image was obtained from (168), Page 2193, Figure 3A, and reproduced with permission conveyed through Copyright Clearance Center, Inc. (C) Low level vacuolation in a Fig4-deficient pale tremor (PLT) mouse. Image was obtained from (169), Page 10, Figure 3, and reproduced with permission from Elsevier. Note the overlap of the large vacuole indicated with the red arrowhead and the nucleus behind it with no indication that the vacuole either compresses the nucleus or causes its compaction. (D) Medium level vacuolation in parietal cortex of mouse inoculated with RML prions. Previously unpublished image produced by the authors. (E) Medium level vacuolation in cerebral cortex of Fig4-deficient PLT mouse. Image obtained from (170), Page 6, Figure 4A, and reproduced with permission conveyed through Copyright Clearance Center, Inc.. (F) High level vacuolation in brainstem of RML prion inoculated mouse. Previously unpublished image produced by the authors. (G) High level vacuolation in neurons of Fig4-deficient PLT mouse. Image obtained from (171), Pages 70, Inset of Figure 2g, and reproduced with permission from Springer Nature. The black arrows shown in Panels 'B' and 'E' were placed by the authors of the respective original publications.

**Figure 6: Cartoon summarizing key components that cause ER and PNS dilation and the proposed relationship of this phenotype to ER hypertrophy and paracrystalline inactive ribosomes.**

The graph depicts a cascade of events that is initiated by the chronic accumulation of toxic prion proteins (top) and ends with neuronal death (bottom) in late-stage prion disease. It can be divided into events that lead to an influx of $Ca^{2+}$ into the cell (depicted on the left-hand side), others that cause activation of the UPR (middle), and chronic stress that lead to a loss of synaptic spines (right-hand side). The graph depicts ER and PNS dilation versus ER hypertrophy as alternative outcomes of cells that attempt to restore normal cell surface levels of functional raft domains (center). We propose that the relative rate of $Ca^{2+}$ build-up in the ER and PNS versus ER biogenesis and sterol synthesis (indicated by triangular shaped nodes) may favor one or the other outcome. In late-stage prion disease, both phenotypes may give rise to paracrystalline assemblies of inactive ribosomes in the cytosol and postsynaptic spines when the supply of tRNAs and mRNAs stalls, because RNAs no longer reach the cytosol and cellular processes have been severed and the cells have died (bottom right). Please see the main text for additional details.

**Supplemental Figure 1: Evidence for co-occurrence of phenotypes of PNS dilation, ER hypertrophy, and paracrystalline aggregates of inactive ribosomes.**

(A) Evidence of the close relationship of PNS dilation and ER hypertrophy in human U2OS cells that were stably transfected with the lamin B receptor (LBR) fused to the yellow fluorescent protein (YFP). Note that the removal or exchange of a single amino acid at the C-terminus of the LBR expression construct converted the phenotype from ER hypertrophy to PNS dilation. Images were obtained from (88), Supplemental Figures S1 and S5, and reproduced with permission of the American Society for Cell Biology, conveyed through Copyright Clearance Center, Inc. (B) Examples of the co-occurrence of PNS dilation and paracrystalline aggregates of inactive ribosomes in the lymphoid cells observed in the hypothermic liver (top panel) or kidney (bottom panel) of chicken herpes virus (Marek's disease). Reproduced from (132), Page 436, Figures 4-7, with permission from Elsevier. (C) Example of co-occurrence of ER hypertrophy and ribosome crystals in hamster kidney cells infected with herpes virus. The image was obtained from (137), Pages 273–274, Figures 4 and 6, and is reproduced here with permission from the American Society for Microbiology (https://journals.asm.org/journal/iai). Except for the schematic of the LBR domain organization, all images in this figure are reproductions from the reports cited. A subset of the images were turned and cropped to emphasize structures relevant for this review. All labels, arrows or boxes, and sizing bars visible in the microscopy images in this figure stem from the respective original publications.

**Table 1: Subset of insults causing massive ER dilation and their links to cellular $Ca^{2+}$ entry**

| Category | Cell death assigned by authors | Insult, compound, or protein | Models of ER and PNS dilation | References | Proposed mechanistic link to ER and PNS dilation phenotype | References |
|---|---|---|---|---|---|---|
| I | ND | Prion disease | Marmosets, mice | (120, 122) | Poisons raft domains and NKAs through accumulation of $PrP^{Sc}$ | (12) |
| | Oncosis | Ouabain | Cats, rats, glia cultures | (12, 22, 79) | Cardiac glycoside known to block ion channel within NKAs | (8, 172) |
| | Oncosis | Palytoxin | Caco-2 | (173) | Converts NKAs to indiscriminate ion leak channels | (174) |
| | Autosis | Hypoxia-Ischemia | Rat brains | (80) | Oxygen restriction causes ATP depletion, thereby inhibiting NKAs | |
| | Autosis | Tat-Beclin 1 | HeLa, U2OS | (80, 81) | Beclin 1 binds to and perturbs NKAs | (81) |
| II | Paraptosis | Tunicamycin | FRO | (83) | Inhibits dolichol pyrophosphate-mediated N-glycosylation in ER, thereby induces UPR. Reduces NKA levels. | (84, 175-177) |
| | Paraptosis | Ophiobolin A | T98G, U373, U343, A172, U251 | (85, 86) | Blocks disulfide formation in ER.<br>Temporary TRPV2 membrane PM targeting can be blocked by tranilast or N-term-TRPV2 overexpression. | (24) |
| | Paraptosis | Cyclosporin A | SiHa, U2OS, HeLa, A549, Saos-2, HaCaT | (87) | Causes UPR by interfering with peptidyl-prolyl isomerization Inhibits ATP1B1-cyclophilin B interaction critical for maturation and trafficking of NKAs. Sensitizes TRPV1 channels to agonists | (87, 93, 178-180) |
| I + III | Paraptosis | ATP1A1, ATP1A2 | U2OS, HeLa | (24) | Causes accumulation of NKA α subunits. Catalytic activity not needed for PNS dilation to manifest. Phenotype can be blocked with tranilast TRPV2 inhibitor. | (24) |
| III | ND | DHCR7s | U2OS | (88) | Catalyzes the last step in cholesterol synthesis pathway. TRPV1 is photosensitized by 7-dehydrocholesterol | (181) |
| | ND | TM7SF2 | U2OS | (88) | Catalyzes the reduction of the C14 unsaturated bond of lanosterol Proposed co-evolution of TRPV channels and cholesterol synthesis pathway | (154) (182) |
| | ND | Truncated LBR | U2OS, HeLa | (88) | Catalyzes the reduction of the C14 unsaturated bond of lanosterol. | (88) |
| IV | Paraptosis | IGF1R-IC | HEK293T, MCF-7, Cos-7 | (89, 183) | Causes TRPV2 calcium leak channel to translocate to plasma membrane. | (96, 97, 184) |
| | Necroptosis | Polyinosinic polycytidylic acid[3] | H29T | (90) | Known agonist of TRPV2 paralog TRPV1 | (185) |
| | Necroptosis | TNF-α [2] | H29T | (90) | TRPV2, but not TRPV4, is upregulated by 10 ng $mL^{-1}$ TNF-α ($P < 0.05$) | (186) |
| V | ND | Aerolysin | BHK | (102, 187-189) | Pore forming toxin causes $Ca^{2+}$ influx by docking to GPI-anchored proteins and forming amphipathic ring-like pore complexes. Induces SREBP. | (102, 187-189) |

**Figure 1**

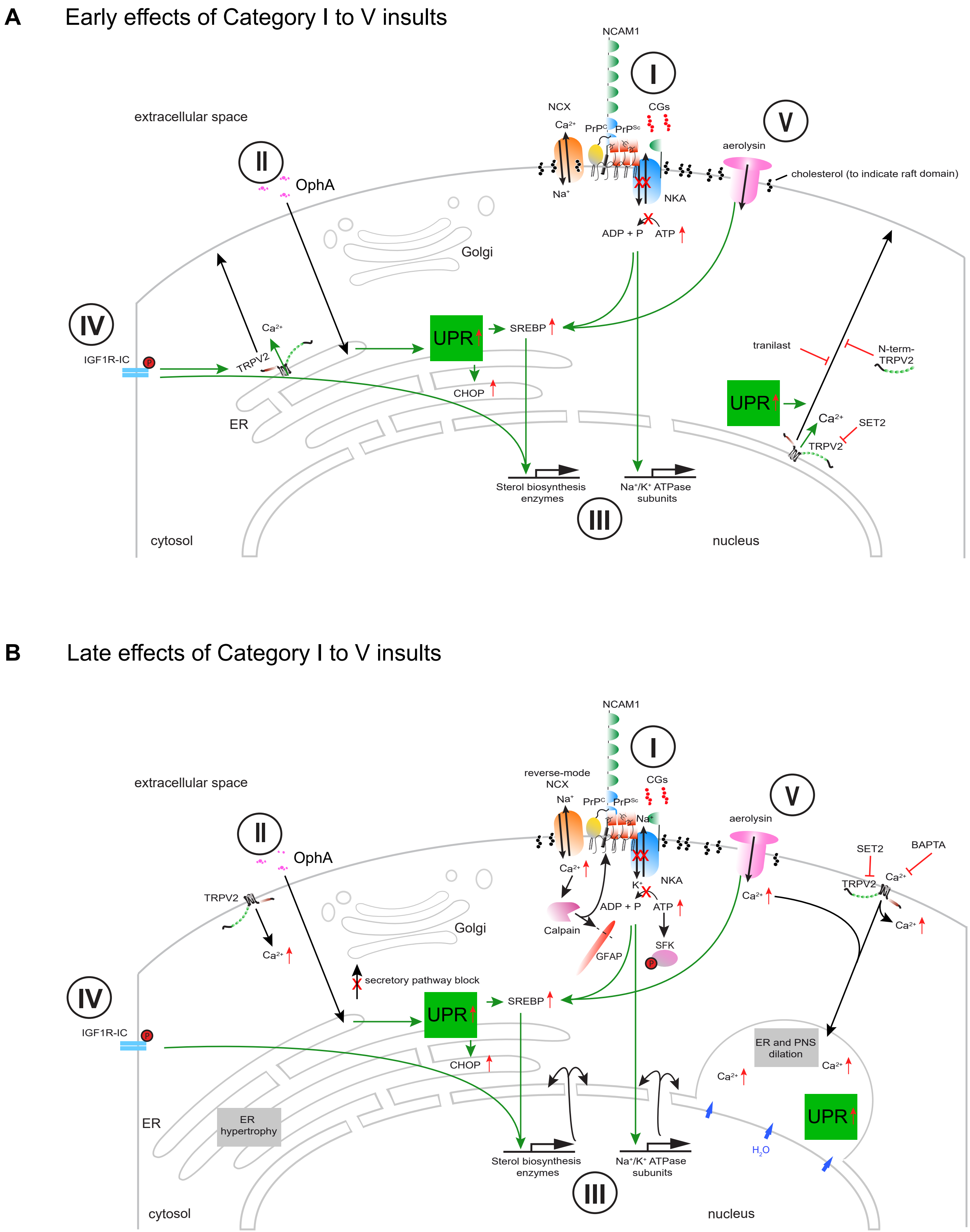
A
Early effects of Category I to V insults
NCAM1
I
NCX
CGs
V
Ca2+
PrPc
PrPSc
aerolysin
extracellular space
cholesterol (to indicate raft domain)
II
OphA
Na+
NKA
ADP + P
ATP
Golgi
IV
Ca2+
UPR
SREBP
tranilast
N-term-TRPV2
IGF1R-IC
TRPV2
CHOP
UPR
ER
Ca2+
SET2
TRPV2
Sterol biosynthesis enzymes
Na+/K+ ATPase subunits
III
cytosol
nucleus
B
Late effects of Category I to V insults
NCAM1
I
reverse-mode NCX
CGs
extracellular space
Na+
PrPc
PrPSc
Na+
V
aerolysin
II
OphA
SET2
BAPTA
Ca2+
K+
NKA
Ca2+
TRPV2
TRPV2
ADP + P
ATP
Ca2+
Ca2+
Calpain
Golgi
SFK
Ca2+
GFAP
secretory pathway block
IV
UPR
SREBP
IGF1R-IC
ER and PNS dilation
CHOP
Ca2+
Ca2+
ER
ER hypertrophy
UPR
H2O
Sterol biosynthesis enzymes
Na+/K+ ATPase subunits
III
cytosol
nucleus

# Figure 2

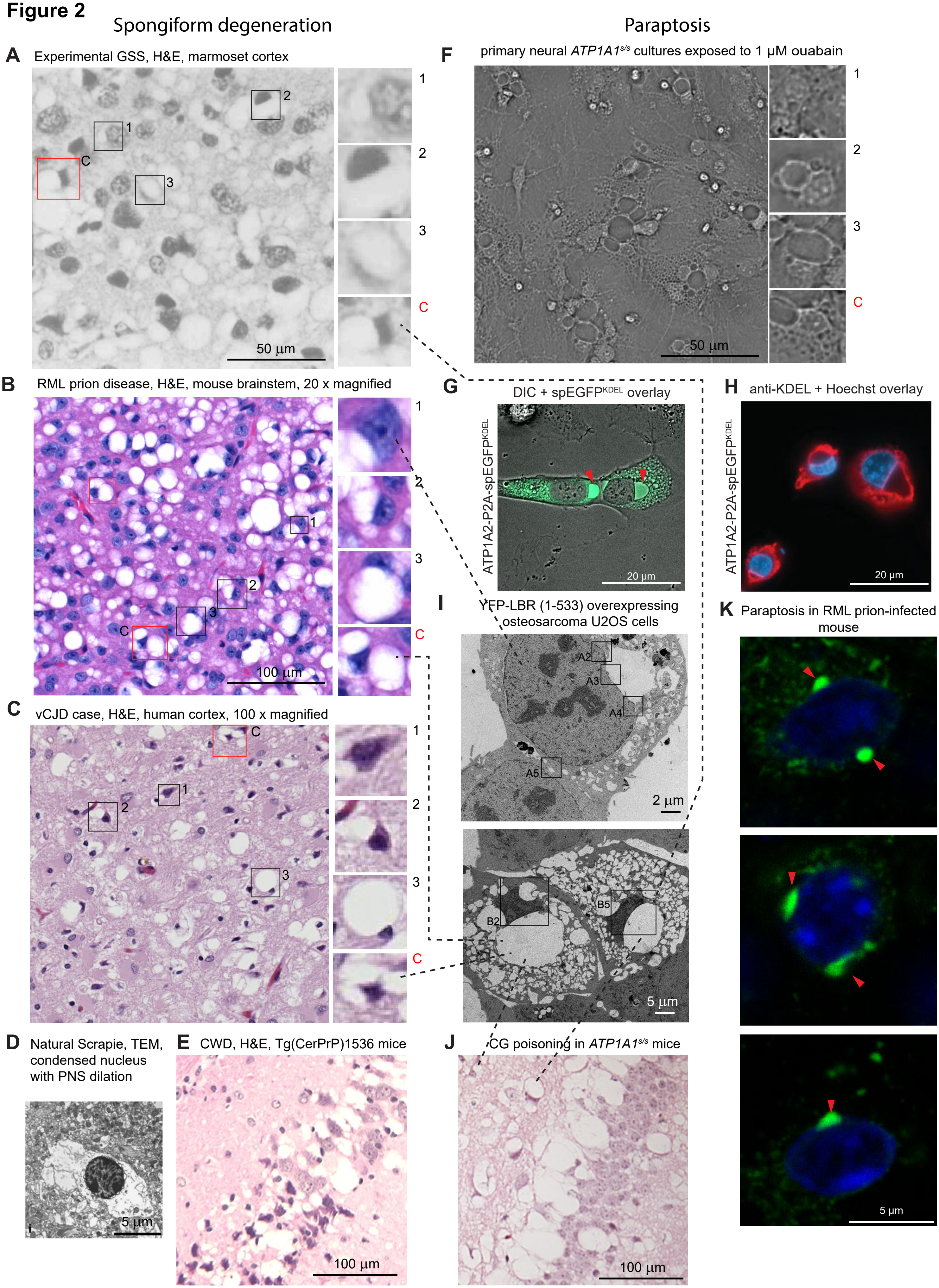

Spongiform degeneration
Paraptosis
A Experimental GSS, H&E, marmoset cortex
50 µm
B RML prion disease, H&E, mouse brainstem, 20 x magnified
100 µm
C vCJD case, H&E, human cortex, 100 x magnified
D Natural Scrapie, TEM, condensed nucleus with PNS dilation
5 µm
E CWD, H&E, Tg(CerPrP)1536 mice
100 µm
F primary neural ATP1A1s/s cultures exposed to 1 µM ouabain
50 µm
G DIC + spEGFPKDEL overlay
ATP1A2-P2A-spEGFPKDEL
20 µm
H anti-KDEL + Hoechst overlay
ATP1A2-P2A-spEGFPKDEL
20 µm
I YFP-LBR (1-533) overexpressing osteosarcoma U2OS cells
A2
A3
A4
A5
2 µm
B2
B5
5 µm
J CG poisoning in ATP1A1s/s mice
100 µm
K Paraptosis in RML prion-infected mouse
5 µm

**Figure 3**

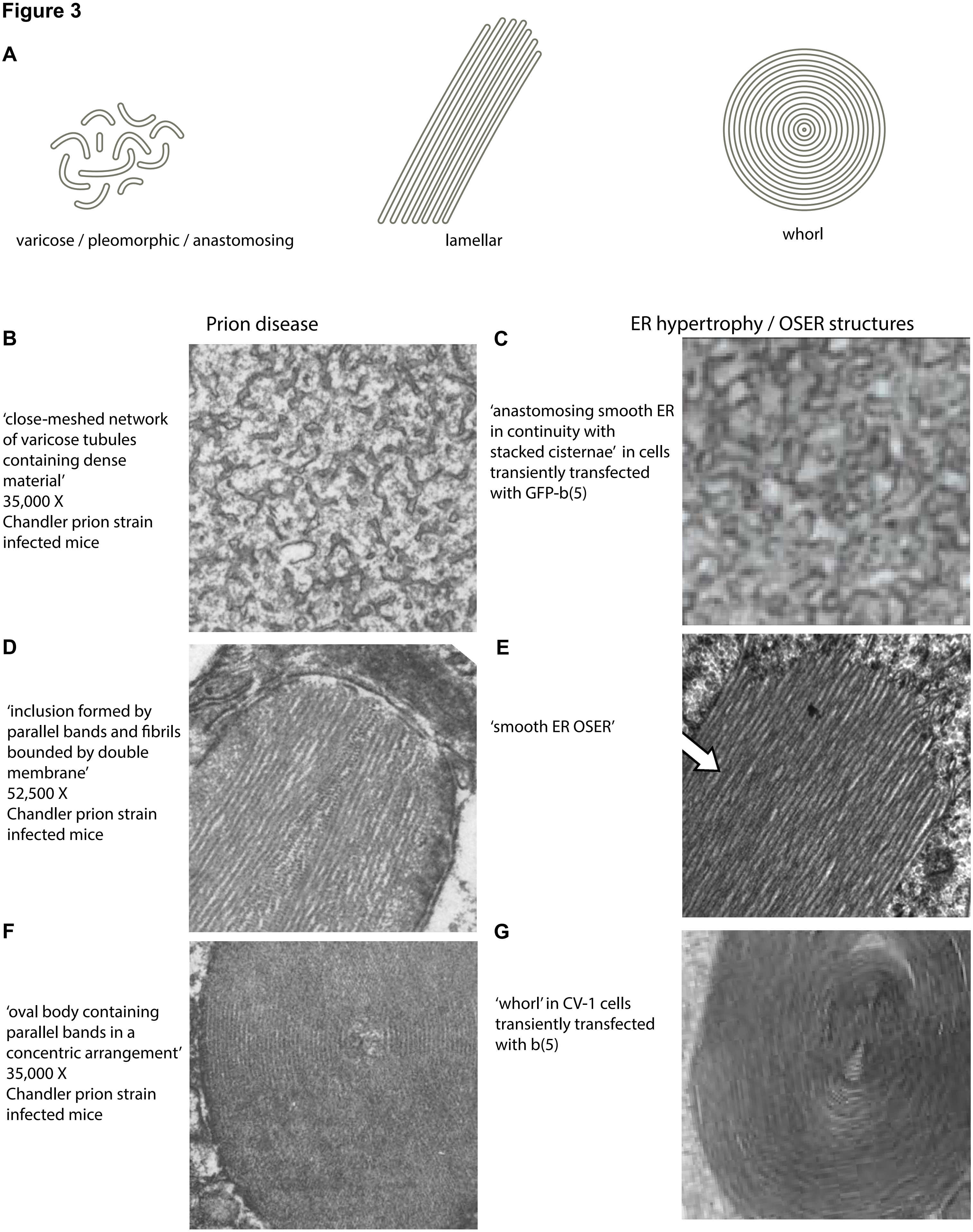
A
varicose / pleomorphic / anastomosing
lamellar
whorl
Prion disease
ER hypertrophy / OSER structures
B
'close-meshed network of varicose tubules containing dense material'
35,000 X
Chandler prion strain infected mice
C
'anastomosing smooth ER in continuity with stacked cisternae' in cells transiently transfected with GFP-b(5)
D
'inclusion formed by parallel bands and fibrils bounded by double membrane'
52,500 X
Chandler prion strain infected mice
E
'smooth ER OSER'
F
'oval body containing parallel bands in a concentric arrangement'
35,000 X
Chandler prion strain infected mice
G
'whorl' in CV-1 cells transiently transfected with b(5)

**Figure 4**

Dense 'virus-like' particles in prion disease

Inactive ribosome aggregates and paracrystals

**A**

'dense particles and rod-like structures'
36,000 X
Chandler prion strain infected mice

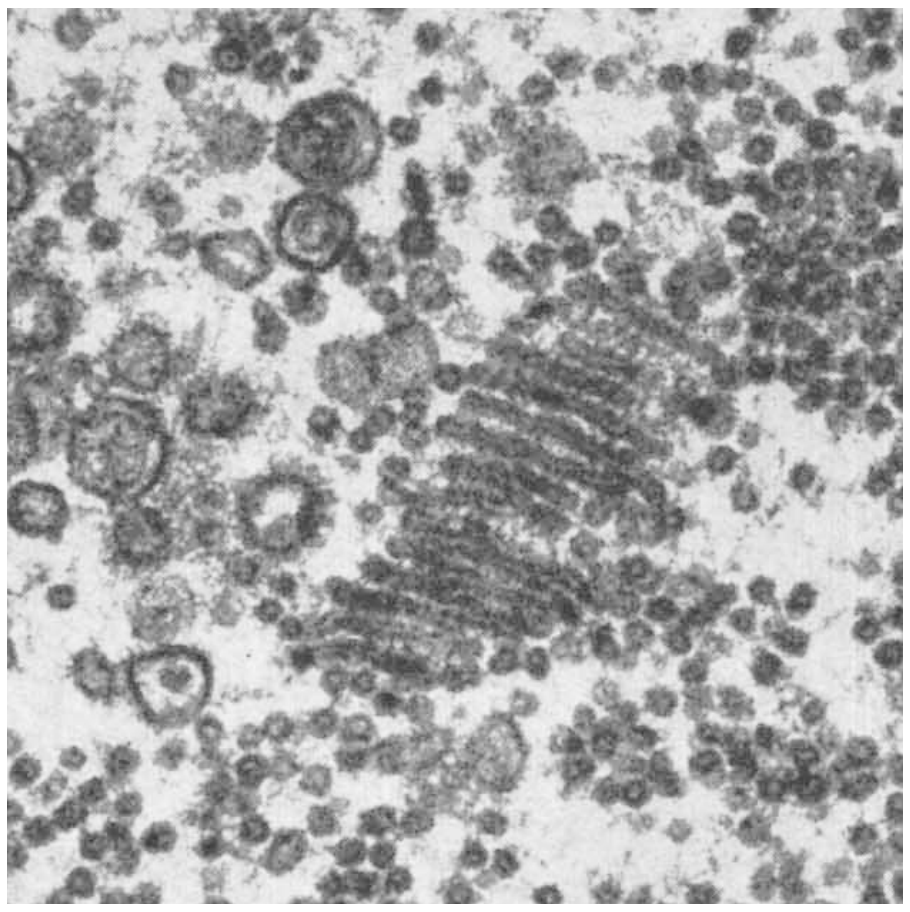

**B**

'solid-line views alternate with dotted line views'
55,000 X
hypothermic chick embryo

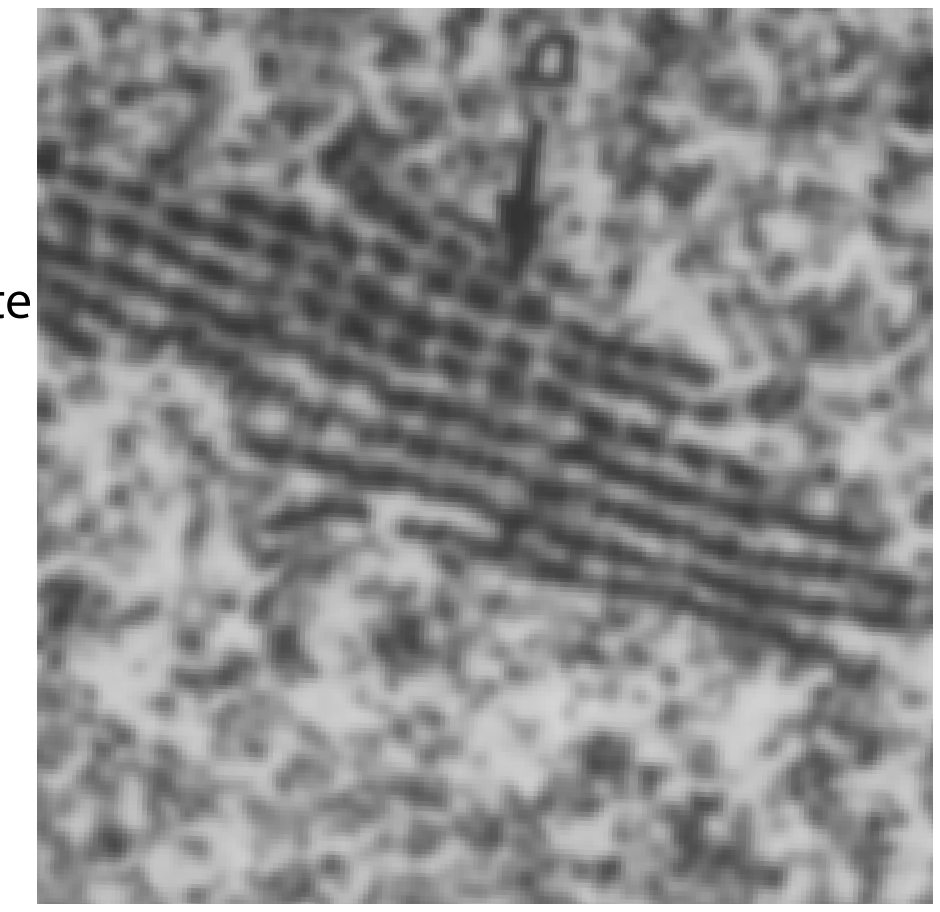

**C**

'process within neuropil contains compact aggregate of dense particles'
'every particle is 23 nm and center-to-center spacing is 37 nm'
36,000 X
Chandler prion strain-infected mice 5 months after inoculation

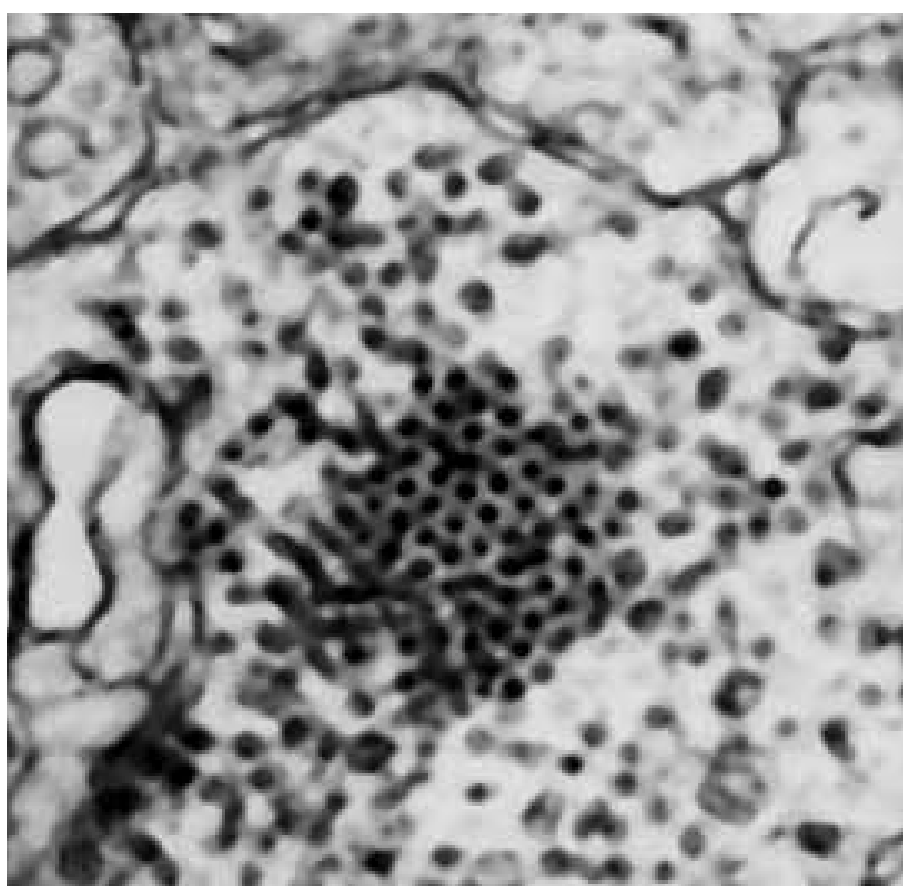

**D**

'ribosome crystals'
25,000 X
hypothermic chick embryo

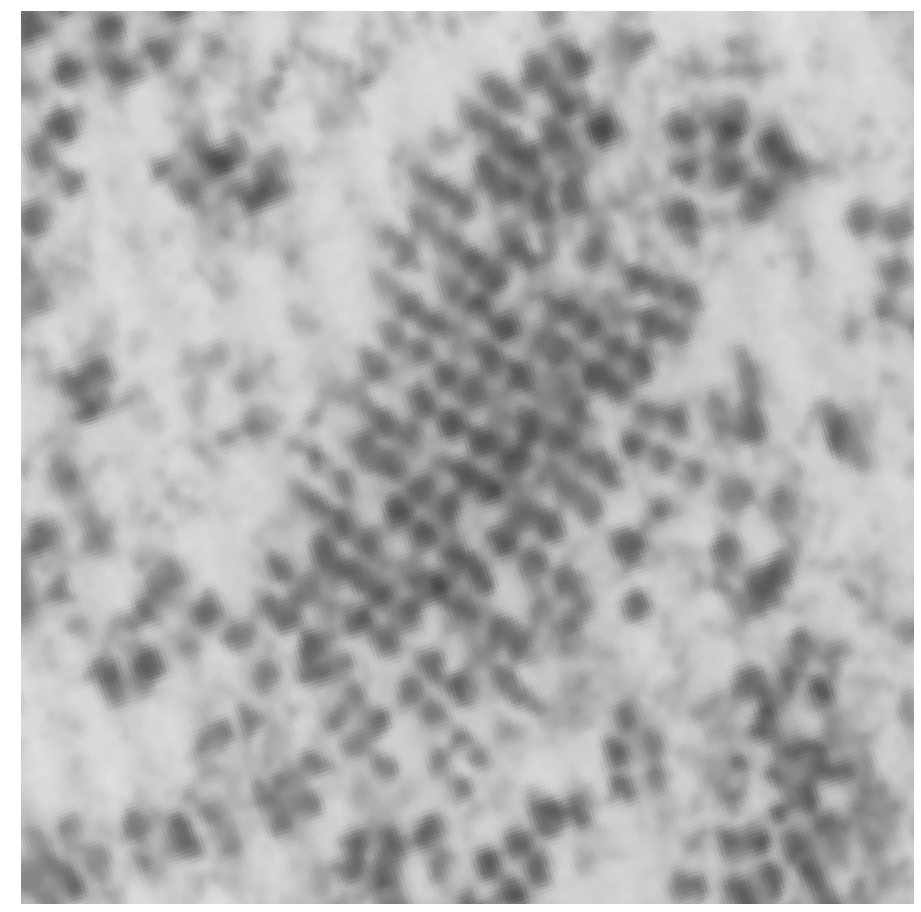

**E**

'30- to 35- nm vesiculo-tubular structures inside a postsynaptic process'
'Klenck' scrapie prion strain-infected mouse

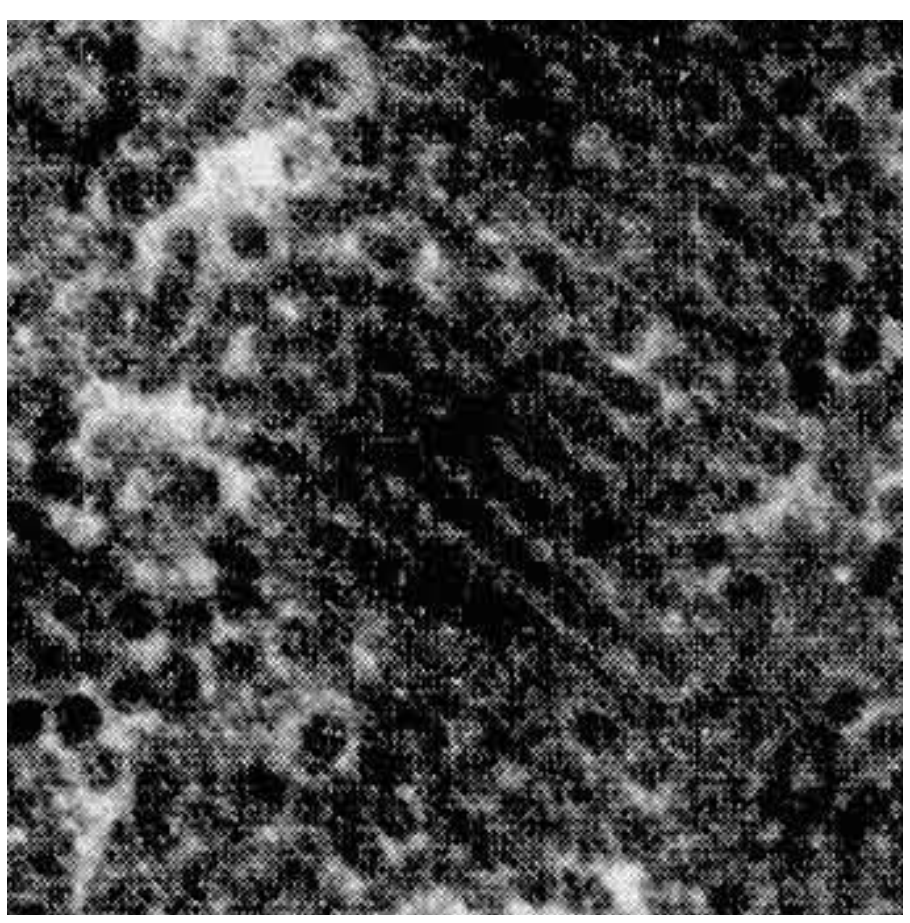

**F**

'ribosome crystals'
25,000 X
hypothermic chick embryo

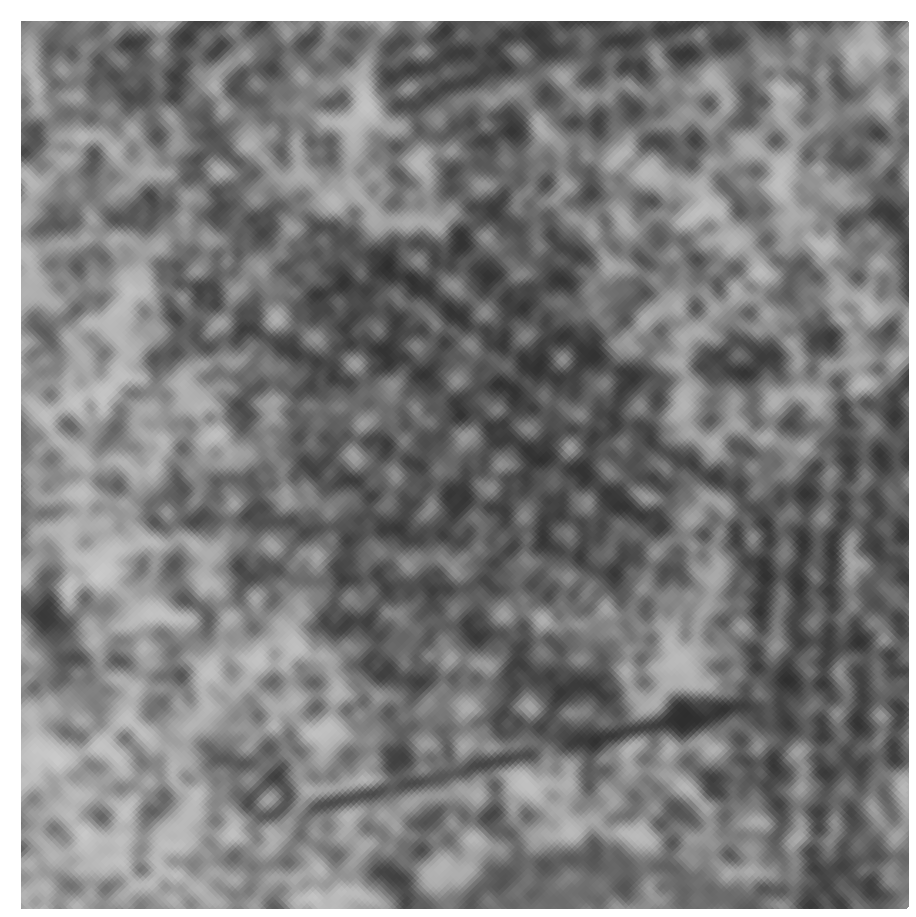

**G**

'post-synaptic element contains particles and vermicular arrays of dense particles'
36,000 X
Chandler prion strain-infected mice

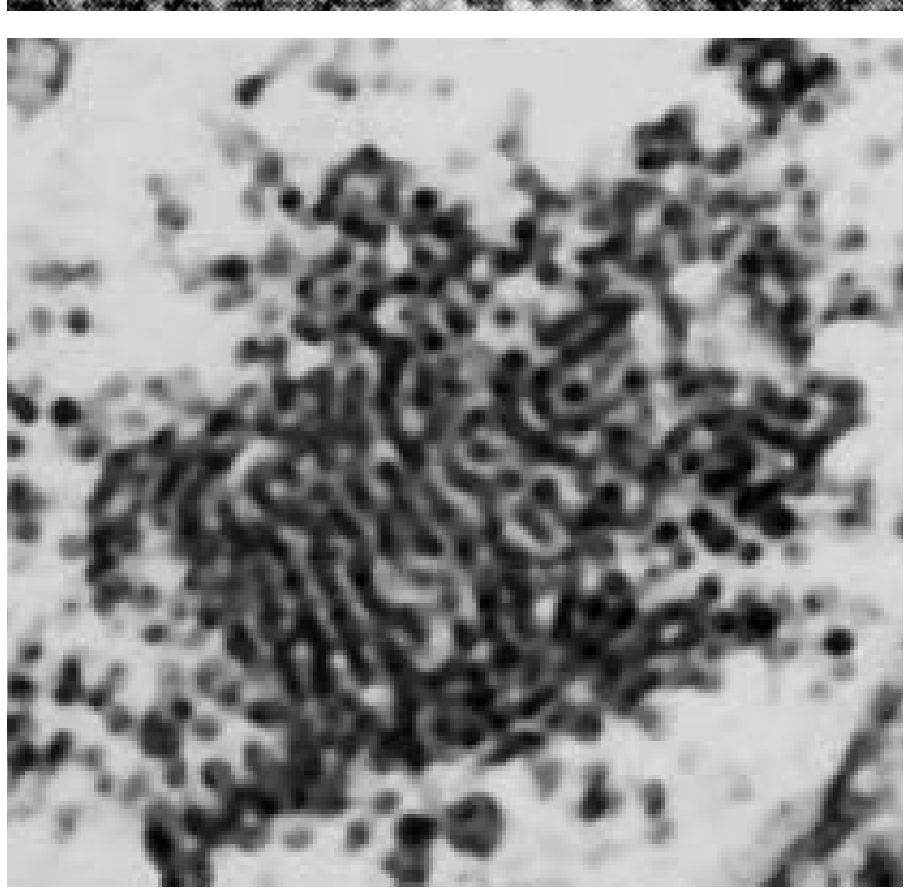

**H**

'polysomes in a cultured follicular epithelial cell'

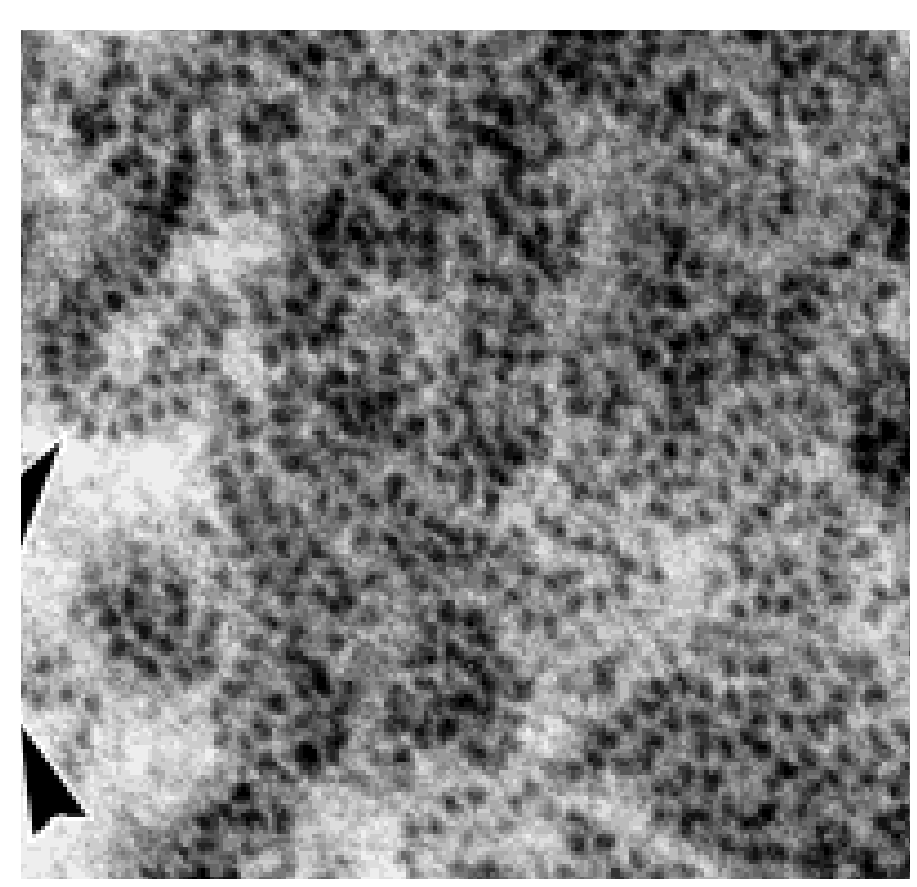

## Figure 5

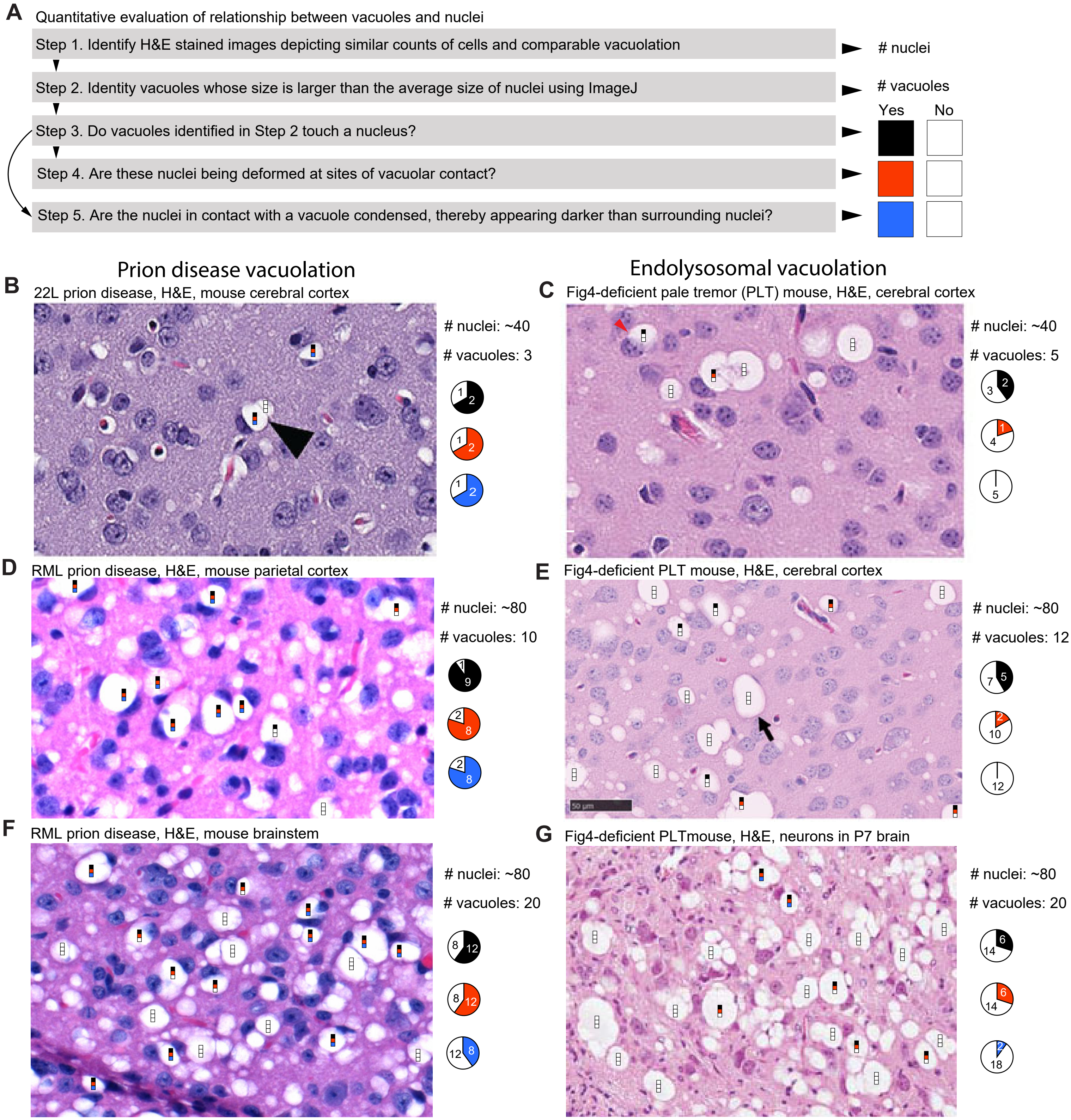
A
Quantitative evaluation of relationship between vacuoles and nuclei
Step 1. Identify H&E stained images depicting similar counts of cells and comparable vacuolation
# nuclei
Step 2. Identity vacuoles whose size is larger than the average size of nuclei using ImageJ
# vacuoles
Yes
No
Step 3. Do vacuoles identified in Step 2 touch a nucleus?
Step 4. Are these nuclei being deformed at sites of vacuolar contact?
Step 5. Are the nuclei in contact with a vacuole condensed, thereby appearing darker than surrounding nuclei?
Prion disease vacuolation
Endolysosomal vacuolation
B
22L prion disease, H&E, mouse cerebral cortex
# nuclei: ~40
# vacuoles: 3
C
Fig4-deficient pale tremor (PLT) mouse, H&E, cerebral cortex
# nuclei: ~40
# vacuoles: 5
D
RML prion disease, H&E, mouse parietal cortex
# nuclei: ~80
# vacuoles: 10
E
Fig4-deficient PLT mouse, H&E, cerebral cortex
# nuclei: ~80
# vacuoles: 12
50 µm
F
RML prion disease, H&E, mouse brainstem
# nuclei: ~80
# vacuoles: 20
G
Fig4-deficient PLTmouse, H&E, neurons in P7 brain
# nuclei: ~80
# vacuoles: 20

**Figure 6**

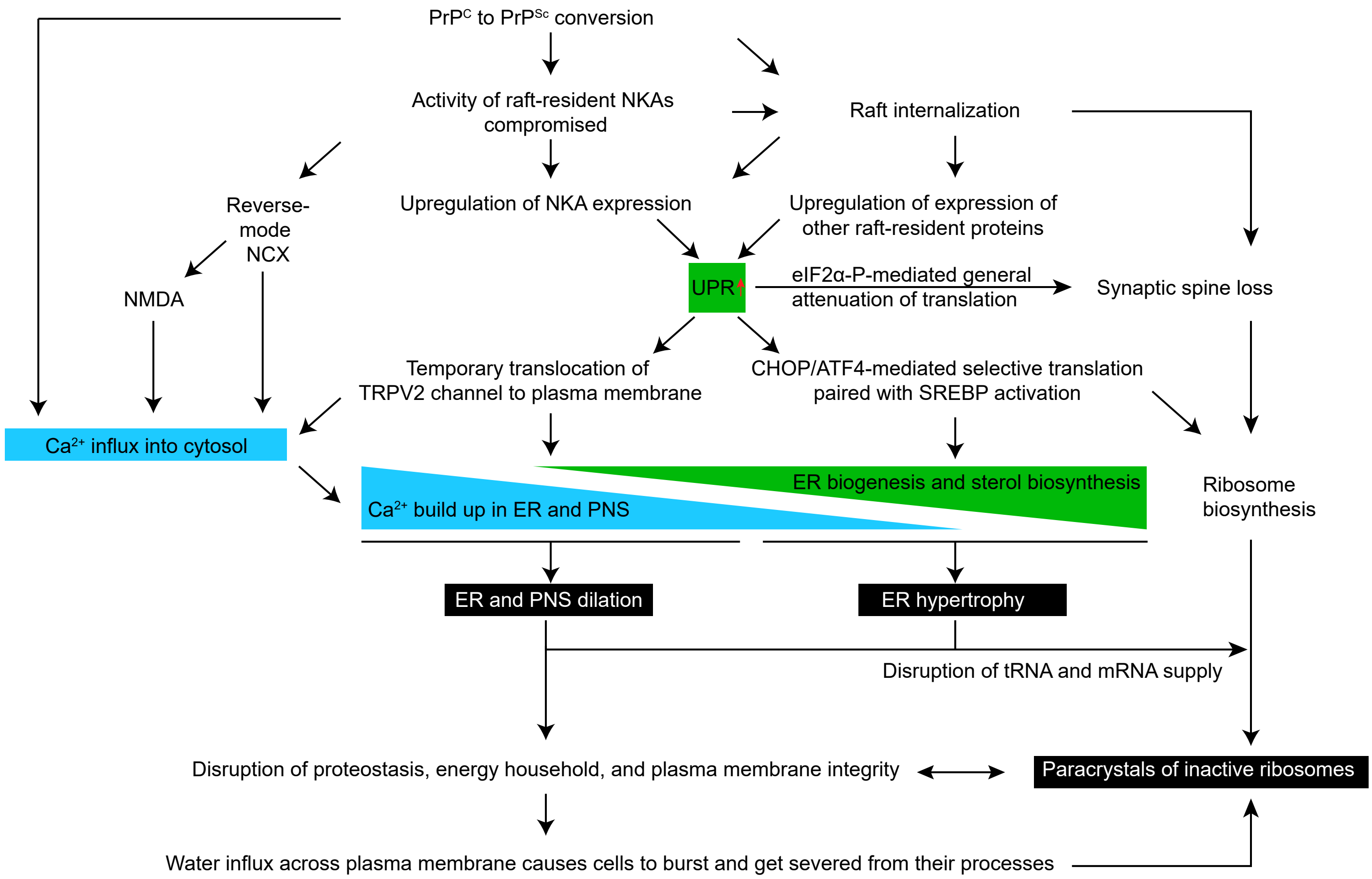
PrP^C to PrP^Sc conversion
Activity of raft-resident NKAs compromised
Raft internalization
Reverse-mode NCX
Upregulation of NKA expression
Upregulation of expression of other raft-resident proteins
NMDA
UPR↑
eIF2α-P-mediated general attenuation of translation
Synaptic spine loss
Temporary translocation of TRPV2 channel to plasma membrane
CHOP/ATF4-mediated selective translation paired with SREBP activation
Ca2+ influx into cytosol
ER biogenesis and sterol biosynthesis
Ca2+ build up in ER and PNS
Ribosome biosynthesis
ER and PNS dilation
ER hypertrophy
Disruption of tRNA and mRNA supply
Disruption of proteostasis, energy household, and plasma membrane integrity
Paracrystals of inactive ribosomes
Water influx across plasma membrane causes cells to burst and get severed from their processes

**Supplemental Figure 1**

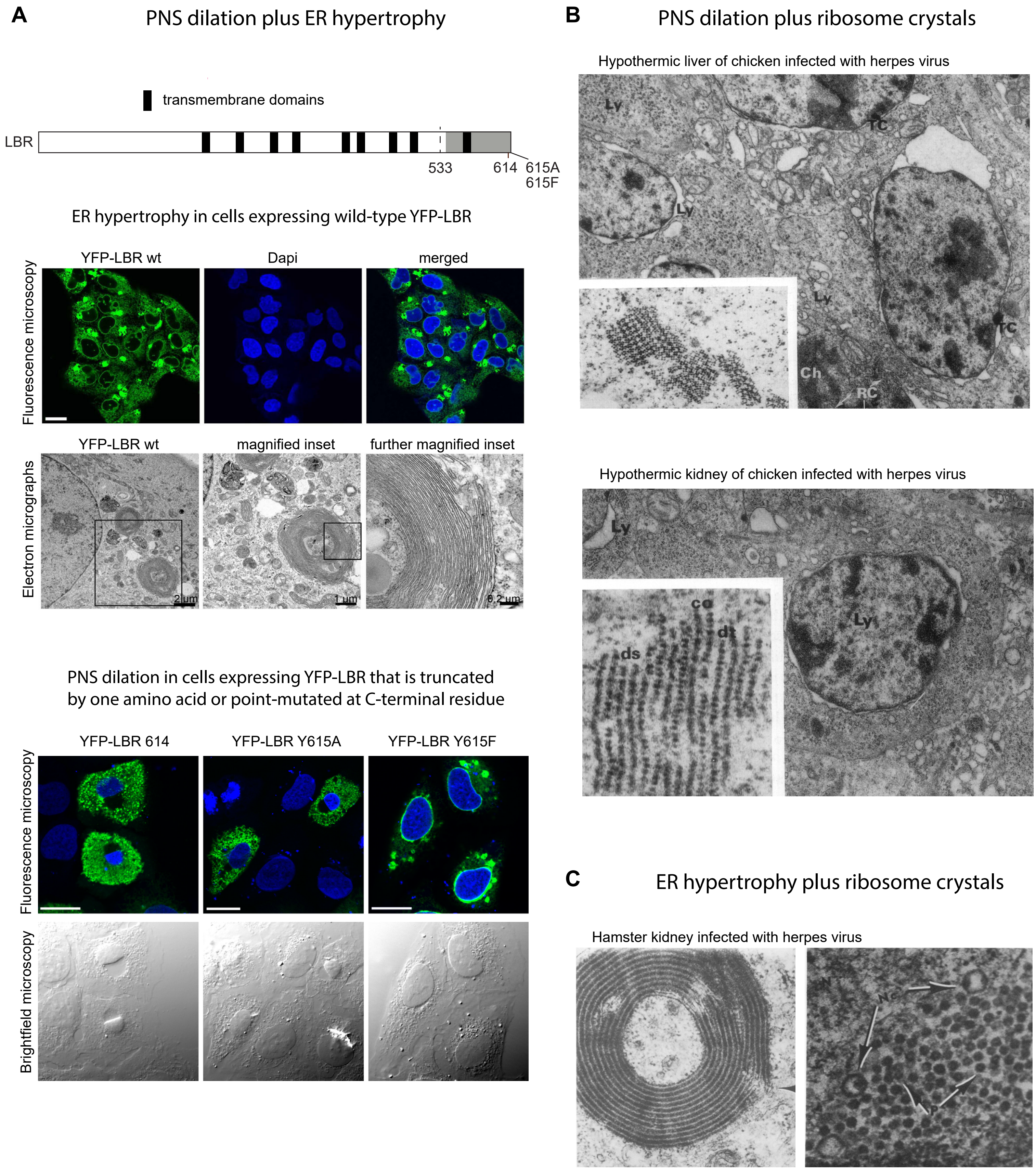